\documentclass[sn-mathphys-num]{sn-jnl}

\usepackage{graphicx}
\usepackage{multirow}
\usepackage{amsmath,amssymb,amsfonts}
\usepackage{amsthm}
\usepackage{mathrsfs}
\usepackage[title]{appendix}
\usepackage{xcolor}
\usepackage{textcomp}
\usepackage{manyfoot}
\usepackage{booktabs}
\usepackage{algorithm}
\usepackage{algorithmicx}
\usepackage{algpseudocode}
\usepackage{listings}
\usepackage{bm}
\usepackage{physics}
\usepackage{float}
\usepackage{mathtools}
\usepackage{subfigure}
\usepackage{epsfig}
\usepackage{color}
\usepackage{tabu}
\usepackage{xspace}

\iftrue

\else

\fi

\newcommand{\casql}{CAS Key Laboratory of Quantum Information, School of Physics, University of Science and Technology of China, Hefei, Anhui, 230026, China}
\newcommand{\casex}{CAS Center For Excellence in Quantum Information and Quantum Physics, University of Science and Technology of China, Hefei, Anhui, 230026, China}
\newcommand{\aihf}{Institute of Artificial Intelligence, Hefei Comprehensive National Science Center, Hefei, Anhui, 230088, China}
\newcommand{\origin}{Origin Quantum Computing, Hefei, Anhui, 230088, China}

\begin{document}

\title{Quantum Computational Insurance and Actuarial Science}

\author[1,2]{\fnm{Huan-Yu} \sur{Liu}}
\equalcont{These authors contributed equally to this work.}

\author[1,2,3]{\fnm{Xi-Ning} \sur{Zhuang}}
\equalcont{These authors contributed equally to this work.}

\author[3]{\fnm{Chao} \sur{Wang}}

\author*[1,2]{\fnm{Qing-Song} \sur{Li}}

\author[3]{\fnm{Meng-Han} \sur{Dou}}

\author[4]{\fnm{Zhao-Yun} \sur{Chen}}

\author*[4]{\fnm{Cheng} \sur{Xue}}\email{xcheng@iai.ustc.edu.cn}

\author*[1,2,4]{\fnm{Yu-Chun} \sur{Wu}}\email{wuyuchun@ustc.edu.cn}

\author*[1,2,4,3]{\fnm{Guo-Ping} \sur{Guo}}\email{gpguo@ustc.edu.cn}

\author[1,2]{\fnm{Guang-Can} \sur{Guo}}

\affil[1]{\casql}
\affil[2]{\casex}
\affil[3]{\origin}
\affil[4]{\aihf}

\abstract{
    In recent years, quantum computation has been rapidly advancing, driving a technological revolution with significant potential across various sectors, particularly in finance. Despite this, the insurance industry, an essential tool for mitigating unforeseen risks and losses, has received limited attention. This paper provides an initial exploration into the realm of quantum computational insurance and actuarial science. After introducing key insurance models and challenges, we discuss quantum algorithms that can address insurance problems based on their mathematical nature. Our study includes experimental and numerical demonstrations of quantum applications in non-life insurance, life insurance, and reinsurance. Additionally, we explore the timeline for quantum insurance, the development of quantum-enhanced insurance products, and the challenges posed by quantum computational advancements. This work systematically constructs the connection between quantum computation and the insurance industry, enhancing the development of insurance while promoting the application of quantum computation to more realistic problems.
}


\maketitle

\section{Introduction}

Insurance is the financial means helping people to avoid potential loss and to hedge against risk from particular sources \cite{ewold1991insurance}.
What lies in the core and crux of insurance practice and actuarial science is to model and evaluate random risky events \cite{vaughan2007fundamentals}.
In the branch of non-life insurance,
The central tasks are to price the premiums paid by an individual policyholder, to model the claims for an insurer to afford when an insured event happens, and to estimate the liability considering the  Incurred But Not Reported (IBNR) and Reported But Not Paid (RBNP) claims \cite{dickson2016insurance}.
Therein, Monte-Carlo simulation, regression, and neural network are introduced to solve complicated statistics mathematics and stochastic process problems \cite{frees2009regression, ohlsson2010non, richman2018ai}.
In contrast, the life insurance branch pays attention to insure life against disability-and-death risk by mortality forecasting \cite{promislow2014fundamentals, lee2000lee, cairns2006two}, 
as well as against live-too-long risk by life annuity pricing \cite{promislow2014fundamentals, brown2001role}.
This can be solved by mortality forecasting models, financial/mortality stochastic processes simulations, and multi-period portfolio optimization \cite{embrechts2022recent, delong2019fair1,delong2019fair2}.
As for the risk management branch, reinsurance and risk indicators are common tools to transfer and quantify the risks, respectively \cite{albrecher2017reinsurance}.
Consequently, stochastic process catching large claims and risk portfolio optimization are both emphasized when studying the reinsurance problem \cite{promislow2014fundamentals,embrechts2022recent,ladoucette2006reinsurance, dickson1996reinsurance}.
Nevertheless, the traditional insurance industry and academia are encountering challenges due to the rapidly growing data size and increasingly complex models, which will require exponentially increased computational resources~\cite{embrechts2022recent}.

 One exciting answer to these problems is quantum computation \cite{Hey2023Book}, which has been widely applied in chemistry \cite{mcardle2020quantum}, biology \cite{outeiral2021prospects, emani2021quantum}, materials \cite{ma2020quantum}, drug design \cite{cao2018potential}, and especially finance \cite{orus2019quantum,egger2020quantum,herman2022survey}. 
Specifically, applying quantum computing methods to insurance problems has also been investigated, including the insurance cash flow modeling \cite{ adam2022potential}, collective risk model~\cite{zhuang2022quantum1}, insurance capital modeling \cite{capital1,capital2}, risk identification \cite{riskide,cber},  insurance claim fraud determination \cite{fraud}, etc. However, quantum computational insurance is still in its infancy compared with other financial branches: 
Firstly, existing works mainly focus on applying quantum computation to specific insurance problems while a systematic study on the general framework has not been established.
Secondly, experimental demonstrations have not been implemented on current quantum computers to pave the way for the applicability and practicality of quantum insurance.
Thirdly, while empirical analysis of these models on real-world data sets is essential for any insurance problems, relevant research is still absent.

We argue that the above phenomena and challenges originate from the significant difference between insurance mathematics and many other financial problems.
From the theoretical aspect, as statistical and actuarial models have been developed and applied to manage risk for hundreds of years, critical analysis and theoretical interpretability have been proven fundamental to understanding their stochastic basis~\cite{beard2013risk}.
The fancy relationships between quantum information and the stochastic nature of risk theory, including multivariate statistical analysis, risk processes, and risks in the collective~\cite{buhlmann2007mathematical}, need to be released.
From the practical aspect, realistic data-based empirical analysis is essential to calibrate and validate insurance and risk models~\cite{mayer1993statistical,wuthrich2013financial,gupta2006model}, while efficient data encoding methods are still challenging for quantum computing, especially in its current Noisy Intermediate Scale Quantum (NISQ) era~\cite{corcoles2019challenges,weigold2020data,zhang2022quantum,lau2022nisq}.

This article provides an early yet insightful exploration of addressing challenges in insurance practice and actuarial science with quantum computation. To provide an intuitive perception of insurance, we first review the problems in the main branches of non-life insurance, life insurance, and risk management. Then we categorize the problems as stochastic modeling tasks, optimization tasks, and machine learning-based tasks, and quantum algorithms to solve these tasks with potential speedup are introduced. Moreover, we also discuss them with respect to the NISQ era \cite{preskill2018quantum} and fault-tolerant quantum computation (FTQC) era \cite{campbell2017roads} based on their requirement for quantum resources, offering a timetable for quantum computational insurance. 

We also numerically and experimentally study several quantum insurance applications corresponding to mainstream insurance branches.
Initially, for the policy excess problem, we designed the quantum excess evaluation algorithm to model the (stochastic) payment and performed numerical simulations. Following this, we experimentally demonstrate solutions for the reinsurance type allocation problems with variational optimization algorithms and the Lee-Carter model \cite{lee2000lee} with machine learning-based algorithms on the superconducting quantum processor ``Wukong" 
\cite{wukong}. Notably, solving the Lee-Carter model for mortality forecasting is the first implementation of an insurance-specific quantum algorithm on a quantum processor using real datasets. These demonstrations and empirical analyses provide further evidence of the feasibility of quantum computational methods for realistic insurance problems.

Beyond the applicability and demonstrability of quantum insurance, we explore its influence and impact on the insurance industry. We consider the three main developmental stages of quantum insurance and highlight how quantum computation is driving the creation of new types of insurance products. Despite these advancements, some fundamental challenges still need to be addressed.

Combining these aspects, our work introduces the connection between quantum computation and the insurance industry. This can pave the way for the development of the insurance industry with quantum computation while promoting the application of quantum computation to more realistic problems.

\section{Problems related to insurance}
\subsection{Non-life Insurance}

Non-life insurance typically covers risks associated with damage to property, legal liabilities, and medical expenses. Specifically:

\paragraph{Collective Risk Model (CRM)}
In CRM \cite{Cossette2019Collective}, the aggregate amount of claims is given by $\sum_{i=1}^{N(t)} Y_i$, where $N(t)$ is the number of claims up to time $t$, modeled by a discrete count process (e.g., a Poisson process), and $Y_i$ is the size of the $i$-th claim, modeled by a continuous stochastic process. A primary concern within CRM is the risk of ruin, which examines the potential for an insurance company to go bankrupt. where the ruin probability is $P_{\text{ruin}}(u) = P(u(t)<0)$, with $u(t)$ the surplus of the insurance company at time $t$ and $u=u(0)$ the initial surplus. In CRM, the arrival of claims is typically assumed to be independent.

\paragraph{Policy Excess and Excess of Loss Reinsurance}
In non-life insurance such as motor vehicle insurance, it is common to consider policies with a Policy Excess \cite{Lourdes2005Dependent}. This means that the risk is divided between the insured party and the insurer according to the contract. Mathematically, if $X$ is the total claim amount and $E$ is the excess amount, the payment by the insurer $I$ can be modeled as $ I = \max(X - E, 0)$. A similar model is used for Excess of Loss Reinsurance, where the risk is shared between the insurer and the reinsurer. The payment by the reinsurer $R$ can be expressed as $R = \max(X-M, 0)$, where $M$ is the retention limit. One common important task in these cases is to model the payments of the insured party, the insurer, and the reinsurer.

\paragraph{Generalized Linear Model}
The generalized linear model \cite{Fahrmeir1992Posterior} is widely used for short-term insurance pricing. The expectation of the outcome $Y$ can be derived as:
\begin{equation}
    \mathbb{E}[Y|X] = \mu = g^{-1}(X\beta),
\end{equation}
where $g$ is the link function and $X\beta$ is the linear predictor. Different link functions and corresponding models are used for different insurance situations. For example, Poisson regression is typically used for modeling the frequency of claims, while Gamma regression may be applied to model the severity of claims.

\paragraph{IBNR}
It often happens that some insurance policies have incurred claims that have not yet been reported \cite{Aiuppa1987Empirical}. The Chain-Ladder Model \cite{Renshaw1998CLM} is a mainstream model used to estimate the IBNR amount. In this model, the claims amount $ A_{i,j} $ is aggregated by accident year $i$ and payment year $j$. The aggregated claims amounts are assumed to be linked by a chain-ladder factor $f_j$ as $ A_{i,j+1} \sim f_j A_{i,j} $ and the core problem is to estimate the chain-ladder factors $f_j$.

\paragraph{Other related Problems}
In addition to the previously discussed issues, non-life insurance also deals with various other problems. These include catastrophe modeling \cite{grossi2005catastrophe}, which involves estimating the impact of large-scale events such as natural disasters; premium calculation, which involves determining the appropriate premium to charge for a given level of risk; reserve setting \cite{8515058}, which entails estimating the reserves needed to cover future claims; and solvency assessment \cite{Stewart1971Assessment}, which evaluates the financial health and solvency of the insurance company. Addressing these challenges often requires the application of advanced statistical and actuarial techniques, alongside robust computational methods.

\subsection{Life Insurance}

Life insurance is a financial product that provides a payout to beneficiaries upon the death of the insured individual. It serves as a crucial tool for financial planning, offering security and peace of mind to policyholders and their families. The valuation of life insurance contracts is a complex and computationally intensive task, involving various factors such as mortality rates, lapse rates, and financial returns.

\paragraph{Valuation of Insurance Contracts}
A key challenge in life insurance is accurately valuing insurance contracts \cite{Kassberger2008valuation}. The expected cash outflow of a whole life insurance policy can be modeled as
\begin{equation}
\text{E}(C) = \sum_{t=1}^{T} q(t) \cdot P_{\text{lapse}}(t),
\end{equation}
where $ \text{E}(C) $ is the expected cash outflow, $ q(t) $  and $P_{\text{lapse}}(t)$ are the payment and the probability of lapse at time $t$. Since policyholders sometimes cancel their contracts, the dynamic lapse rate $ \lambda(t) $ is considered:
\begin{equation}
P_{\text{lapse}}(t) = 1 - e^{-\lambda(t)  t}.
\end{equation}
These equations help capture the complexities involved in contract valuation by integrating both the timing of payments and the likelihood of policy termination over time.

\paragraph{Mortality Forecasting} Multivariate statistical analysis and data science techniques are introduced to solve the mortality forecasting problems, which have many variants. For instance, the Lee-Carter model \cite{lee2000lee} is a popular approach for mortality forecasting:
\begin{equation}\label{leecarter}
\ln m_{x,t} = \alpha_x + \beta_x \kappa_t + \epsilon_{x,t},
\end{equation}
where $ m_{x,t} $ is the mortality rate for age $ x $ at time $ t $, $ \alpha_x = (\sum_{x=1}^T m_{x,t} )/T $ represents the age-specific component, $ \beta_x $ captures the sensitivity to the time-varying index $ \kappa_t $, and $ \epsilon_{x,t} $ is the error term. This model and its variants are extensively discussed in \cite{richman2018ai}.

\paragraph{Multi-period Optimization}
For life insurance, especially for life annuities, Multi-period Optimization \cite{Brodt1983MIN} is an essential long-term financial management instrument. This involves optimizing the portfolio over multiple periods, taking into account not only the return and variance of each assessment but also the transaction costs. The objective function can be formulated as
\begin{equation}
\max \left\{ \sum_{t=1}^{T} \left( R_t - \frac{1}{2} \gamma \sigma_t^2 - C_t \right) \right\},
\end{equation}
where $ R_t $ is the return at time $ t $, $ \gamma $ is the risk aversion coefficient, $ \sigma_t^2 $ is the variance of the return, and $ C_t $ represents the transaction costs. 

\paragraph{Other Related Problems}
In addition to valuation, mortality forecasting, and multi-period optimization, there are several other related problems of interest in the field of life insurance. First, underwriting risk \cite{Brockett1982Underwriting} refers to the assessment of risk associated with insuring a new policyholder, and it often involves statistical and machine learning techniques to predict the likelihood of a claim. Additionally, longevity risk \cite{Waegenaere2010Longevity} is another crucial concern, as it represents the risk that policyholders live longer than expected, which can significantly impact the financial stability of life insurance companies. This issue is particularly relevant for annuity products. Furthermore, ensuring that life insurance products comply with regulatory standards is essential, as the standards can vary significantly across different jurisdictions. Finally, understanding and predicting customer behavior \cite{ervasti2013understanding}, such as lapse rates and claim frequencies, is critical for better managing and pricing insurance products.

\subsection{Risk Management}

Risk management is a critical aspect of financial and insurance industries, aiming to identify, assess, and prioritize risks followed by coordinated efforts to minimize, monitor, and control the probability or impact of unfortunate events. Effective risk management ensures the stability and profitability of financial institutions.

\paragraph{(Conditional) Value at Risk}
The Value at Risk (VaR) and conditional Value at Risk (cVaR) are common risk measures in risk management \cite{duffie1997overview}. VaR is defined as the minimum loss bound at a given risk level $\alpha \in [0, 1]$, such that the realized loss $x$ is less than this bound with a probability of at least $\alpha$. Mathematically, it is expressed as
\begin{equation}
\text{VaR}_\alpha[X] = \inf{\{x|\mathbb{P}[X \le x] \ge \alpha \}}.
\end{equation}
On the other hand, cVaR is the expected loss conditioned on the loss being under the bound $VaR_\alpha[X]$ defined as:
\begin{equation}
\text{cVaR}_\alpha[X] = \mathbb{E}[X|0 \le X \le \text{VaR}_\alpha[X]].
\end{equation}
Managing the total risk of an insurance portfolio is an essential problem. For each kind of insurance contract, one must decide whether to transfer the risk to the private market or to self-manage it within the insurance pool by allocating its reinsurance type.

\paragraph{Risk Transfer and Reinsurance}
In the context of insurance, risk transfer and reinsurance are crucial strategies \cite{albrecher2017reinsurance}. Risk transfer involves moving the risk from the insurance company to another party, typically through reinsurance contracts. Reinsurance helps insurers to manage their risk exposure and protect against significant losses. The decision on the type and extent of reinsurance involves evaluating the trade-offs between the cost of reinsurance and the benefits of reduced risk exposure.

\paragraph{Portfolio Risk Optimization}
Portfolio risk optimization \cite{pfaff2016financial} is another important aspect of risk management. This involves optimizing the allocation of assets and liabilities to minimize risk while maximizing returns. Techniques such as mean-variance optimization and stochastic programming are commonly used. The objective function for portfolio optimization can be formulated as
\begin{equation}
\max \left\{ \sum_{i=1}^{n} \left( R_i - \frac{1}{2} \gamma \sigma_i^2 \right) \right\},
\end{equation}
where $ R_i $ is the return of asset $ i $, $ \gamma $ is the risk aversion coefficient, and $ \sigma_i^2 $ is the variance of the return of asset $ i $.

\paragraph{Other Related Problems}
There are several other related problems in risk management. First, credit risk management \cite{Witzany2017Credit} involves assessing the risk of loss due to a borrower's failure to make payments. This often includes the use of credit scoring models and default probability estimations. Second, operational risk management \cite{moosa2007operational} focuses on risks arising from internal processes, systems, and external events, including fraud and cybersecurity threats. Furthermore, liquidity risk management \cite{Holmström2000Liquidity} ensures that the institution can meet its short-term obligations without incurring significant losses.

\section{General framework for quantum computational insurance}

In this section, we establish a general framework for quantum computational insurance.
After a brief introduction to the basics of quantum computation, we explore the connections between insurance problems and quantum computation.
Firstly, the fundamental concept of probability distributions can be efficiently encoded on quantum states by state preparation algorithms.
Secondly, the essential model of risk processes can be simulated and analyzed by quantum stochastic simulation algorithms.
Thirdly, the more complicated collective and portfolio of risk processes can be optimized by quantum optimization algorithms.
Finally, the rapidly growing data-driven insurance models can be trained by quantum machine learning models.
Following them, more specific algorithm implementations are left to the next section.

\subsection{Basics in quantum computation}

This part briefly reviews some basic concepts in quantum computation. The first is the quantum bit (qubit), the basic information in quantum computation. It is different from the classical bit as it can be in a superposition of states, a two-dimensional vector in the Hilbert space:
\begin{equation}
    |\psi\rangle = \alpha|0\rangle + \beta|1\rangle,
\end{equation}
where $|0\rangle$ and $|1\rangle$ are the two basis states, $\alpha$ and $\beta$ are complex amplitudes satisfying $|\alpha|^2+|\beta|^2=1$. The multi-qubit state is within the tensor product of individual qubit space, indicating that an $n$-qubit state represents a $2^n$-dimensional vector in the Hilbert space.

In quantum computation, we perform operations on qubits to accomplish specific tasks. The basic operations in the quantum circuit model are quantum logic gates, which are mathematically described by unitary matrices. Specifically, the single-qubit and two-qubit gates are commonly used. For instance, the single-qubit $X$ gate is the NOT gate:
\begin{equation}
    X = \begin{pmatrix}
        0 & 1 \\
        1 & 0
    \end{pmatrix} \to \begin{cases}
        X|0\rangle =& |1\rangle, \\
        X|1\rangle =& |0\rangle.
    \end{cases}
\end{equation}
The two-qubit Controlled-X gate applies the $X$ gate on the target qubit according to the state of the control qubit:
\begin{equation}
    CX = \begin{pmatrix}
        1 & 0 & 0 & 0 \\
        0 & 1 & 0 & 0 \\
        0 & 0 & 0 & 1 \\
        0 & 0 & 1 & 0
    \end{pmatrix} \to \begin{cases}
        CX |00\rangle =& |00\rangle, \\
        CX |01\rangle =& |01\rangle, \\
        CX |10\rangle =& |11\rangle, \\
        CX |11\rangle =& |10\rangle.
    \end{cases}
\end{equation}
An $n$-qubit quantum logic gate can process the $2^n$-dimensional information in one step, outperforming classical logic gates. It can also cause entanglements or interference between qubits. These features make quantum computation potential to achieving speedups over classical computations. Quantum algorithms based on these principles have been designed for various regions, including chemistry, biology, finance, etc.

\subsection{Quantum algorithms for insurance}

\begin{figure}
    \centering
    \includegraphics[width=0.8\linewidth]{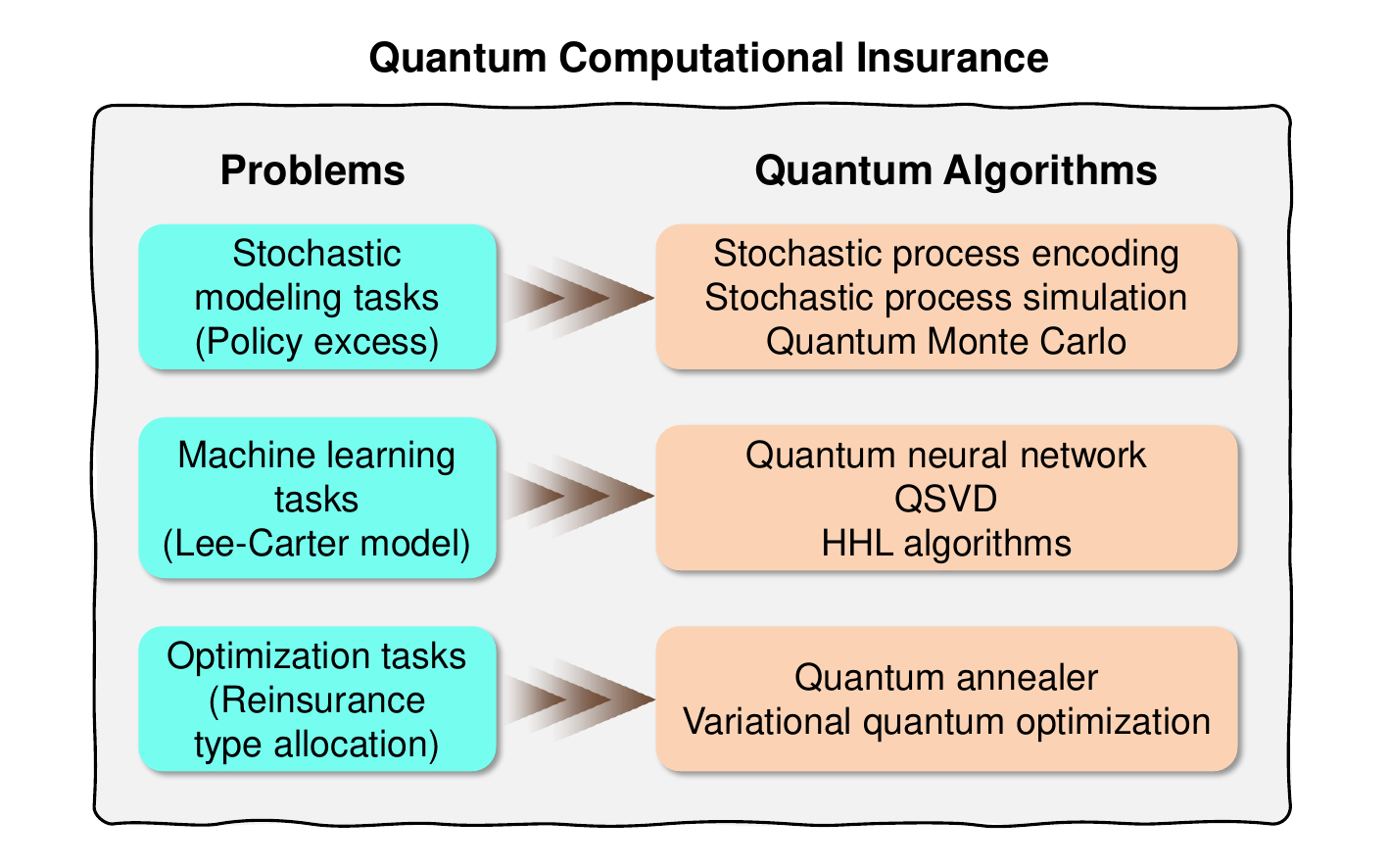}
    \caption{Quantum algorithms for insurance problems, which are categorized as stochastic modeling tasks, machine learning tasks, and optimization tasks. The policy excess problem, the Lee-Carter model, and the reinsurance type allocation problem are typical examples in these classes and are demonstrated numerically and experimentally in the next section.}
    \label{fig:algreview}
\end{figure}

This part discusses quantum algorithms for insurance problems. An overview is shown in Fig. \ref{fig:algreview}.

As aforementioned, modeling stochastic events is the core task of insurance and actuarial science. For instance, CRM and the policy excess problem introduced before. Quantum stochastic modeling involves embedding classical stochastic information into a quantum computer, simulating the stochastic process, and extracting the associated classical information for practical utilization. Efficiency in these steps is required for quantum advantages. Fortunately, quantum stochastic modeling could reduce the space exponentially with a further quadratic speedup. Firstly, statistical distributions, as well as initial states of stochastic processes, can be prepared in a quantum superposition, wherein the intrinsic symmetry of the underlying distribution can be employed to reduce the time complexity and mitigate the classical quantum input bottleneck.
Secondly, for a stochastic process, the Markov dynamics described by a transition matrix can be simulated with quantum operations, herein the memory-less property can be employed to extend to continuous-time processes at a low cost.
While the non-Markovian behaviors are related to an open quantum system, whose simulation has been widely studied \cite{rmpWeimer2021Simulation, delgado2024quantum}. Finally, to evaluate the desired information from the expectation of moments, we implement the quantum-enhanced Monte Carlo integration (QMC). It applies amplitude estimation to the quantum samplings described above with a quadratic accelerate \cite{montanaro2015quantum,layden2023quantum}.

In spite of stochastic modeling, many practical insurance problems, such as insurance pricing and mortality forecasting, can be solved by data-driven methods of machine learning \cite{richman2018ai}. And quantum machine learning would be an appealing choice. In the NISQ era, quantum machine learning is realized via the well-known variational quantum algorithms \cite{Cerezo2021Variational}, which are hybrid quantum-classical that combine parameterized quantum circuits as neural networks and classical computers as optimizers. Parameterized quantum circuits are set to be shallow to mitigate the hardware noise.

While in the FTQC era, where quantum circuits and the qubit number are greatly improved, it is possible to implement quantum machine learning algorithms based on linear algebra. The core subroutine in these algorithms is the solution of linear equations using the widely applied HHL algorithm \cite{hhl, Duan2020survey}. For instance, using quantum computers, The vision transformer \cite{Han2023vision}, which is successfully applied on natural language processing tasks, can achieve speedups \cite{xue2024endtoendquantumvisiontransformer}. The singular value decomposition (SVD) to extracting features can also be accelerated with quantum algorithms \cite{Bravo2020quantum}. 

When considering life annuity and risk assessment management, optimization plays an essential role. Quadratic Unconstrained Binary Optimization (QUBO) problems can be translated into finding the ground state of Ising models, and a quantum annealer can derive the minima \cite{albash2018adiabatic, rosenberg2015solving,venturelli2019reverse}. Variational optimization algorithms like the quantum approximation optimization algorithms \cite{QAOA} can also be applied for these goals.

\section{Results}

In this part, we numerically and experimentally study insurance problems covering the main branches as non-life insurance, life insurance, and reinsurance. 
Based on the discussion of quantum algorithms, we respectively solve them with quantum stochastic modeling algorithms, quantum optimization algorithms, and quantum machine learning-based algorithms. 
First, we develop a quantum algorithm to model the (stochastic) payment in the excess evaluation problem.
The quantum computing framework QPanda \cite{qpanda} is used for this task. Then we solve the reinsurance type allocation problem with variational quantum optimization algorithms and the Lee-Carter model with quantum singular value decomposition algorithms on ``Wukong" \cite{wukong}, which is a high-performance superconducting quantum computer with the average relaxation time $T_1 = 14.84 \mu s$ and $T_2=1.85 \mu s$. The average fidelity of single-qubit gates and two-qubit gates are respectively $99.7\%$ and $96.3\%$.

\subsection{Quantum excess evaluation}\label{subsec:excess}

In this part, we numerically solve the policy excess problem, which aims to model the payment for the reinsurer to pay $\mathbb{E}[  \max (0, X-M)  ]$. In this task, we consider the loss following the distribution 
\begin{equation}
f(x,\mu,\sigma) = \frac{1}{x\sigma\sqrt{2\pi}} e^{- \frac{ (\ln x -\mu)^2  }{2\sigma^2}  },
\end{equation}
where $\mu=0$ and $\sigma=1$ are used. The threshold $M(x)$ is set to be a non-decreasing function
\begin{equation}
    M(x) = \begin{cases}
        0, & x\leq 1, \\
        0.6(x-1)+1, & x>1.
    \end{cases}
\end{equation}
In this case, the payment can be re-expressed as $R_{\text{theory}}=\int_1^\infty f(x) (x-M(x))dx$. Due to the convergence properties of the integral, it can be truncated with a certain level of precision. In this simulation, the excess $x$ is varied from 0 to 10 and the corresponding truncated payment is $R_{\text{truncate}} = \int_{x=1}^{10} g(x) (x-M(x)) dx$, where $g(x) = \frac{f(x)}{\int_{x=0}^{10} f(x)  dx}$. In real simulations, we also need to discrete the range into $N=2^n$ points (claims), The loss of each claim is $X_j,0\leq j\leq N-1$. Then the payment becomes $R_{\text{discrete}} = \sum_{X_j\geq 1} f(X_j) (X_j-M(X_j))$. 

We develop a quantum algorithm for this problem, which uses $n+2$ qubits according to the discrete of the excess range. The quantum algorithm starts with preparing the quantum state representing the distribution of loss, then a quantum subtracter is applied to evaluate each $X_j-M(X_j)$ in parallel. Then a sequence of controlled rotations is applied to establish the connection between the payment and the results when measuring auxiliary qubits. Details of the algorithm can be found in Appendix \ref{qeea}. 

\begin{figure}
    \centering
    \includegraphics[width=0.6\linewidth]{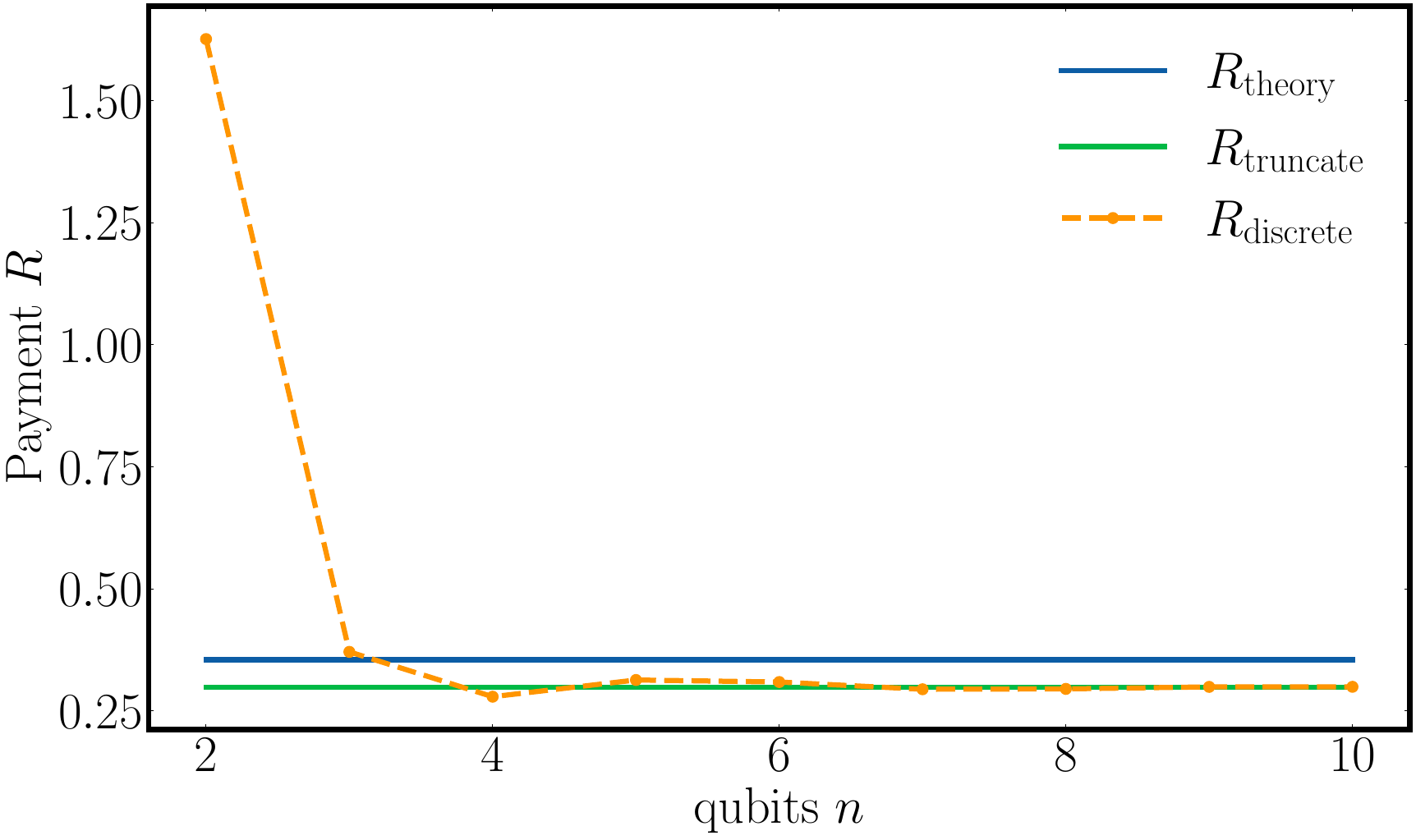}
    \caption{
    \textbf{Simulation results of excess evaluation.}
    $R_{\text{theory}}$ and $R_{\text{truncate}}$ represent the theoretical and the truncated one, demonstrating that only considering $x\in[0,10]$ provides adequate accuracy. $R_{\text{discrete}}$ is the payment evaluated with our algorithm, showing that as the number of qubits increases, the accuracy is improved and converged to $R_{\text{truncate}}$
    }
    \label{fig:qeea}
\end{figure}

The numerical simulation result of this task is shown in Fig. \ref{fig:qeea}. Several conclusions can be observed. Firstly, it works when we truncate for large $x$. Secondly, the correctness of the algorithm grows with the number of qubits. Thirdly, the algorithm can characterize the excess evaluation problem well.

\subsection{Quantum reinsurance type allocation}\label{subsec:rei}

As aforementioned, insurance companies pay reinsurance premiums for risk diversification and long-term stability
with different risk characterizations and charges for the following two reinsurance types:
Self-managed insurance pooling across different regions can reduce the insurance portfolio variance with lower premiums while being fragile to wide-ranged catastrophic \cite{mahul2010government, xu2010systemic}.
Private insurance purchasing from the international reinsurance market can further diversify risk at the cost of higher reinsurance premiums \cite{miranda1997systemic}.
It is an essential yet intractable problem to allocate the reinsurance type for hundreds of insurance products to minimize the variance of the risk portfolio \cite{porth2015portfolio}.

More formally, given the retained risk $R_j$ in the self-managed reinsurance pool that has not been transferred to the private insurance market, one aims to minimize the total variance of the total risk $R_{\text{total}}=\sum_j x_jR_j$ with varying reinsurance type $x_j \in \{0, 1\}$ as
\begin{equation}\label{eq:var}
    \min_{\textbf{x}\in \{0, 1\}^n} \text{Var}  [R_{\text{total}}] \qquad \text{s.t.} \quad \sum_{j=1}^n x_j = pn.
\end{equation}
Herein the risk $R_j$ can be either a vector of historical data or a random variable from stochastic simulation, and the proportion $p$ satisfying $0<p<1$ represents the average allocation probability and is assumed to be fixed (for example, $p=0.5$ as in \cite{porth2015portfolio}).

Expand the above variance, we have:
\begin{equation}
    \text{Var}  [R_{\text{total}}] = \sum_j \text{Var}[x_jR_j] + \sum_{j\neq k} \text{Cov}[x_jR_j,x_kR_k],
\end{equation}
where $\text{Cov}[X,Y]$ is the covariance between $X$ and $Y$. Using $\text{Var}[X]=\text{Cov}[X,X]$, the equation becomes:
\begin{equation}
    \text{Var}  [R_{\text{total}}] = \sum_{j k} \text{Cov}[x_jR_j,x_kR_k] = \sum_{jk} x_jx_k \text{Cov}[R_j,R_k].
\end{equation}
Denote the covariance matrix $V$ with elements $v_{jk}= \text{Cov}[R_j,R_k]$, We have:
\begin{equation}
    \text{Var}  [R_{\text{total}}] = \sum_{jk} x_jv_{jk}x_k = x^TVx.
\end{equation}

In this task, we randomly generate the covariance matrix, and a VQA is applied. We generate the ansatz with some parameterized unitary $|x(\bm{\theta})\rangle = U(\bm{\theta})|0\rangle$. To evaluate the variance with quantum measurements, we apply the map $x_j\to (I-Z_j)/2$ such that $x_j=\langle x_j |(I-Z_j)/2|x_j\rangle,x_j=0,1$. Then the variance becomes:
\begin{equation}
    \text{Var}  [R_{\text{total}}] =\langle x|H_{\text{cost}}|x\rangle, \qquad  H_{\text{cost}} = \sum_{j,k=1}^n v_{jk}\frac{I-Z_j}{2}\frac{I-Z_k}{2},
\end{equation}
Then it is seen that evaluating the variance becomes a measure of the expectation value of the Hamiltonian $H_{\text{cost}}$. Finding $x$ satisfying $\sum_jx_j=pn$ that minimizes the variance becomes finding the eigenstate corresponding to the smallest eigenvalue of the Hamiltonian's specific subspace. Finally, we can apply a classical optimizer to optimize the results.

\begin{figure*}[ht]
    \centering
    \includegraphics[width=0.9\linewidth]{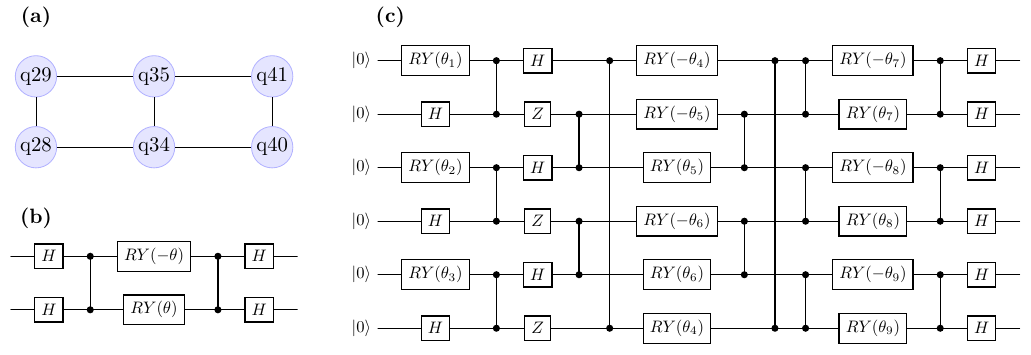}
    \caption{
    \textbf{Setup for the reinsurance type allocation problem.}
    (a) The 6 superconducting qubits used in the quantum computer ``Wukong".
    (b) The quantum circuit for the RBS gates.
    (c) The quantum circuit for the ansatz, with 6 qubits and 3 RBS layers, after some simplifications.
    }
    \label{fig-rbs}
\end{figure*}

We applied 6 superconducting qubits, illustrated in Fig. \ref{fig-rbs}(a), for this task. To satisfy the constraint, instead of a traditional penalty function method, we utilize a more efficient problem-specified ansatz consisting of two parts:
a parameterized initialization layer to prepare a quantum state satisfying the constraint, and repeating layers of reconfigurable Beam splitter (RBS) gates to transform the state while keeping the constraint unviolated \cite{kerenidis2021classical, johri2021nearest,rbsrev}. The RBS gate is $\text{diag}\left\{1, \begin{pmatrix}
    \cos\theta &-\sin\theta \\
    \sin\theta & \cos\theta
\end{pmatrix},1 \right\}$, which interacts between $|01\rangle$ and $|10\rangle$ while keeps $|00\rangle$ and $|11\rangle$ unchanged. A quantum circuit implementation of this gate is shown in Fig. \ref{fig-rbs}(b). The number of RBS layers is set to be 3 and the whole ansatz after some simplifications is shown in Fig. \ref{fig-rbs}(c).

While the theoretical output maintains the constraints, any deviation due to hardware noise is corrected by excluding erroneous shots from measurements. We apply a measurement-based error mitigation strategy to combat quantum hardware noise.
It first runs several quantum circuits to benchmark the probability of flips when measuring 0 and 1. Then after the measurement outcome is obtained, we perform a reverse operation to get the noise-mitigated result.

\begin{figure}
    \centering
    \includegraphics[width=0.6\linewidth]{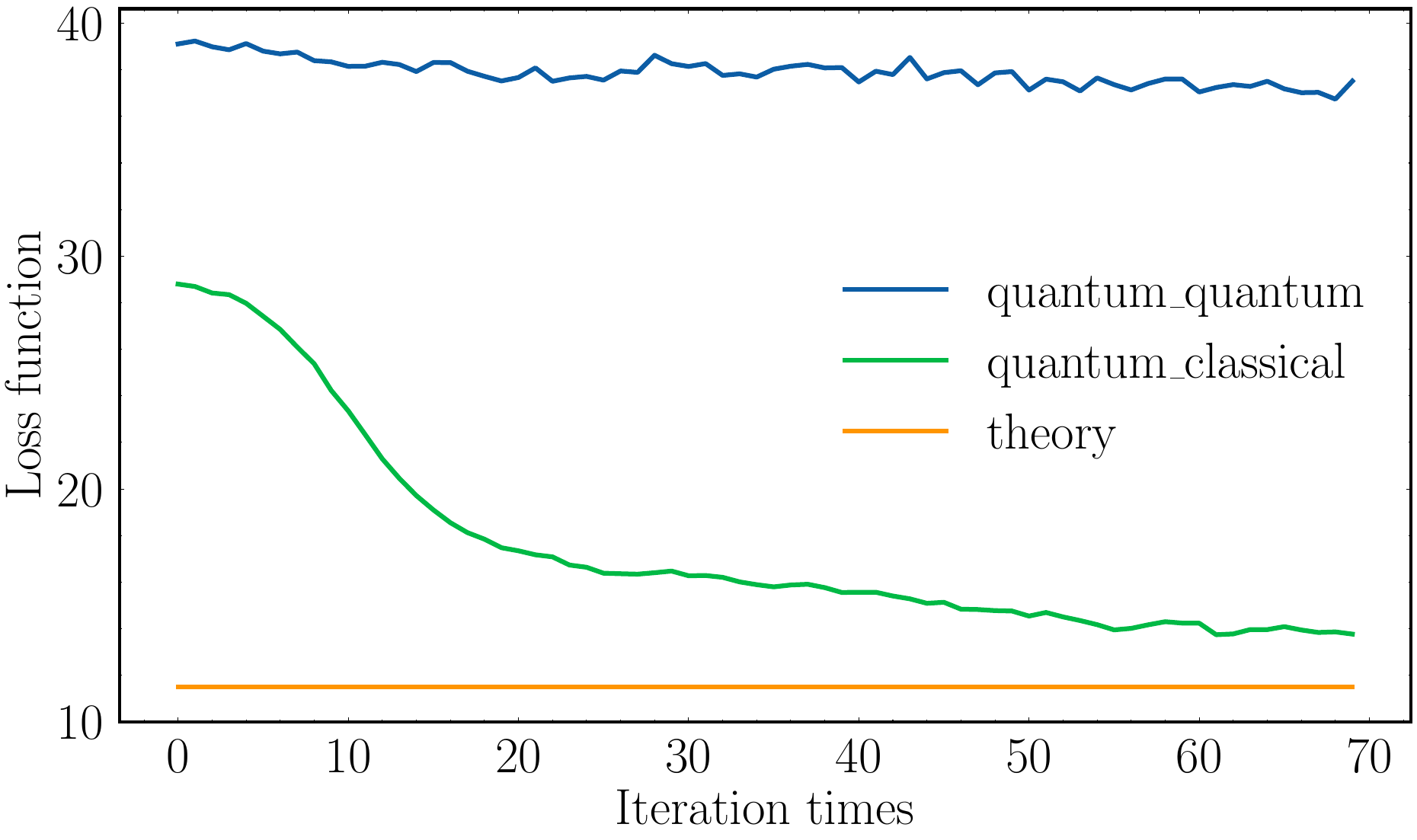}
    \caption{\textbf{Experimental results for the reinsurance type allocation problem.} 
    The blue line illustrates the decrease in the loss function as outputted by quantum computers. The green line indicates the classically computed loss function using the same parameters. The orange line represents the target value.}
    \label{fig-qita}
\end{figure}

The result is shown in Fig. \ref{fig-qita}. The bottom yellow line indicates the minimum orange of the loss function. We optimize the parameters based on the output of quantum computers while using the parameter to compute the noiseless loss function classically, which are labeled by ``quantum\_quantum" and ``quantum\_classical", respectively. We can see that as the optimization goes, the loss function is decreased, and the corresponding ``classical" loss function is closer to the target value, which is capable of generating the target $x$.

\subsection{Quantum Lee-Carter solver}
Here, we report experimental results for solving the Lee-Carter model described in Eq. (\ref{leecarter}). Given a list of data $m_{x,t}$, the core task is to estimate $\beta_x$ and $\kappa_t$. This is usually done by first constructing the matrix $D$ whose elements are $D_{x,t}=\ln m_{x,t}-\alpha_x$. Then a singular value decomposition is performed on $D$, and the left(right) singular vector related to the maximum singular value is associated with $\beta_x(\kappa_t)$. Then these data can be used for further predictions. We apply real-world data obtained from \footnote{\url{https://www.mortality.org}}. The year $t$ is from $2014$ to $2017$ and the age $x$ is from $0$ to $10-14$. Then the matrix size of $D$ is set as $4\times 4$.

We apply the quantum singular value decomposition algorithm \cite{Bravo2020quantum}. A brief introduction to this algorithm is shown in Appendix \ref{method:qsvd}. A total of 4 superconducting qubits are used, and a total of 200 optimization steps are evaluated. Similar to the previous experiment, we directly optimized the parameters on a quantum computer and used those parameters for benchmarks. Denote the optimized left and right vectors in the optimization process as $\beta$ and $\kappa$, then to characterize the optimization process, we observe the Frobenius norm distance between the data matrix $D$ and our constructed matrix $\beta\kappa$, $\Vert D-\beta\kappa \Vert_F$, and the  Kullback-Leibler (K-L) divergence between the optimized vector and the target one, $D_{\text{KL}}(\beta\Vert\beta_x)$ and $D_{\text{KL}}(\kappa\Vert\kappa_t)$, where $D_{\text{KL}}(p\Vert q)=\sum_i p_i \log (p_i/q_i)$.

\begin{figure}[h]
    \centering
    \includegraphics[width=0.8\linewidth]{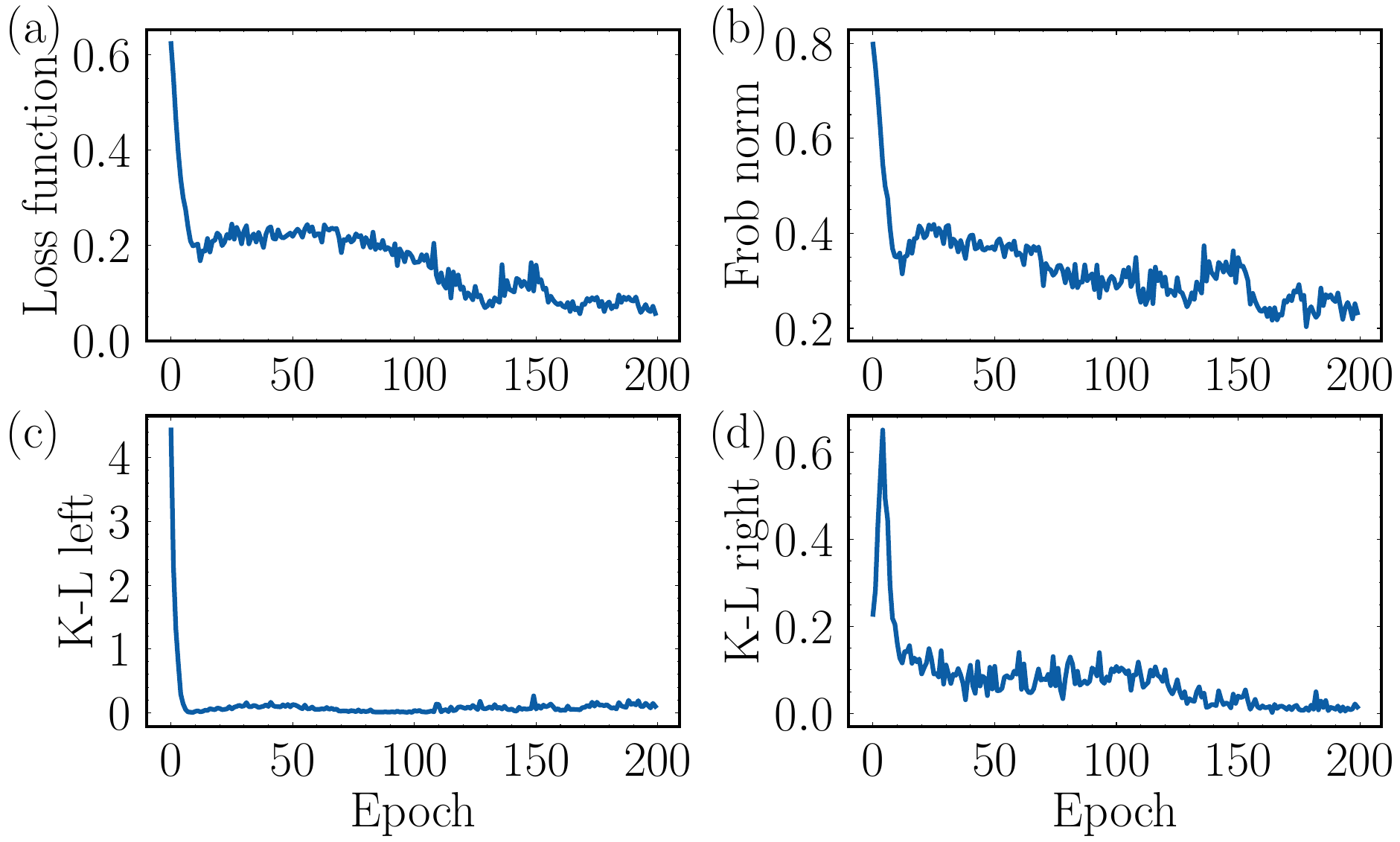}
    \caption{
    \textbf{Experimental results of the Lee-Carter model}.
    (a) Evolution of the loss function in the optimization process.
    (b) Evolution of the  Frobenius norm distance between the data matrix $D$ and our constructed matrix $\beta\kappa$, where $\beta$ and $\kappa$ are the left and right vectors in the optimization process.
    (c) Evolution of the K-L divergence between the theoretical vector $\beta_x$ and $\beta$.
    (d) Evolution of the K-L divergence between the theoretical vector $\kappa_t$ and $\kappa$.
    }
    \label{fig:qsvd}
\end{figure}

The experimental results are shown in Fig. \ref{fig:qsvd}. It shows that the loss function decreases with the optimization steps. During this process, both the Frobenius norm distance between the data matrix and our constructed one and the K-L divergence between the theoretical singular vector and our estimated singular vector also decrease, indicating the successful performance of the target singular value decomposition. Note that this is the first time an insurance-specific real-world data problem is experimentally solved with real quantum computers.

\section{Conclusion and discussion}

In the preceding sections, we delved into the realm of quantum computational insurance and actuarial science. We examined critical issues across the principal branches of insurance, including non-life insurance, life insurance, and reinsurance. Following this, we explored the application of quantum algorithms to address these challenges. Additionally, we conducted both numerical and experimental studies on several insurance problems using quantum algorithms. Besides the introduction of the general paths to applying quantum computing to insurance, these demonstrations clarified the associated technical details.
The connection between quantum computation and the insurance industry is established and enhanced, which can promote the development of both quantum computing and insurance.

In this section, we outline the roadmap for the future of quantum computational insurance. In the NISQ era, the available qubit number of quantum operation depth is limited due to the noise \cite{preskill2018quantum}.
Although experiments have shown early quantum advantages on some specific applications \cite{arute2019quantum}, there is still a long way to universal quantum computing.
While one of the most challenging problems is to strive for fault-tolerant quantum computing,
how can the insurance industry make use of quantum computation in the NISQ era?
We believe that both theoretical and experimental breakthroughs focusing on quantum insurance and actuarial science will be achieved. 
On the aspect of theoretical development, many quantum insurance algorithms will be proposed and proved to have quantum speed-up.
Quantum stochastic modeling will be applied to study risk and ruin theory with many applications as suggested in the previous section.
Also, quantum machine learning and optimization will be further developed on prediction and portfolio optimization problems.
As for the aspect of experiments, quantum insurance algorithms, including but not limited to variational quantum algorithms and quantum Monte Carlo simulations, can be performed on quantum processors.
The insurance industry can help to find the best scenarios and applications of quantum insurance and bring more attention to this cross-field.
With the cooperation of quantum computation scientists, the insurance industry, and academia, quantum algorithms driven or inspired by insurance can reveal more power of quantum computation and further applicability to insurance.

In the second stage of fault-tolerant quantum computing, errors on quantum processors can be corrected \cite{Gambetta_2017, Campbell_2017}.
These high-quality logical qubits, together with larger qubit numbers, will enable us to consider and perform more realistic quantum insurance applications \cite{Campbell_2017}.
As the power of quantum computation has been released, many exciting insurance applications and products can come true.
The risk can be estimated and evaluated more precisely, and such an evaluation process can be even online and personalized, based on quantum computation with Internet and Internet of Things data.
The potential loss can be predicted more early and precisely, benefiting from quantum computational climate predicting, mortality forecasting, and fat-tail events modeling, to name but a few.
Furthermore, quantum natural language processing can help to accelerate complicated tasks such as AI customer service and consultation \cite{coecke2020foundations},   and quantum pattern recognition can speed up claim analysis  \cite{trugenberger2002quantum}.
Nevertheless, one should also notice that, besides those quantum benefits mentioned above, there are quantum threats as well.
Impacted by universal quantum computation, there will be lots of unexpected and consecutive influences that can be viewed as a source of uncertainty and risks.
We believe that one of the most challenging problems is the modeling and evaluating of such a new source of risks.
Once again, we caution that the insurance industry and quantum scientists should cooperate together to handle these challenges.

In this work, we have discussed the major ideas and impacts of quantum computational insurance and actuarial science, and have considered most of the possible applications and challenges at present and in the future.
We also emphasized that we are far away from the final answer to those fundamental problems of quantum computational insurance and actuarial science.
The applications and potential challenges within the context of quantum insurance should be invested in detail and considered carefully and the communication and cooperation of the insurance industry and quantum scientists are necessary.

Similar to other practical applications, quantum algorithms still face significant challenges. For instance, encoding a general distribution into a quantum computer requires the preparation of the corresponding quantum state, which demands exponential resources \cite{qsp1,qsp2}. Although $n$ qubits can encode $2^n$ amplitudes, each shot yields only a single $n$-bit binary string after performing the desired operations, thereby greatly limiting efficiency. Additionally, noise in quantum computers remains a challenge for most quantum algorithms. To address these issues, researchers are exploring several solutions. Quantum Random Access Memory \cite{Giovannetti2008qram} and advanced encoding strategies \cite{sipqc} can be utilized for efficient state preparation. Advanced sampling methods, such as shadow tomography \cite{st,st2}, offer faster readout. Furthermore, quantum error mitigation \cite{qem} strategies are being developed to combat noise in quantum hardware.

\bmhead{Acknowledgements}
This work has been supported by the National Key Research and Development Program of China (Grant No. 2023YFB4502500), the National Natural Science Foundation of China (Grant No. 12404564), and the Aeronautical Science Foundation of China (Grant No. 2022Z073004001).

\bmhead{Data Availability}
The data that support the findings of this study are available from the corresponding author upon reasonable request.

\begin{appendices}

\section{Quantum excess evaluation algorithm}\label{qeea}

\begin{figure*}[ht]
    \centering
    \includegraphics[width=0.8\linewidth]{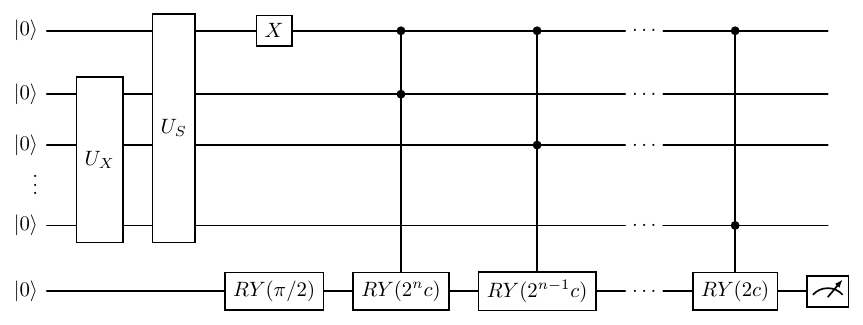}
    \caption{\textbf{Quantum circuits for the quantum excess evaluation algorithm}. The unitary operations $U_X$ and $U_S$ are used to prepare the state $|\psi_1\rangle$ and realize the subtraction operations. Followed controlling operations rotate the measurement qubit. The desired value is associated with the probability of obtaining $|1\rangle$ in multiple shots.}
    \label{fig:circuitqee}
\end{figure*}

Here, we provide details about the quantum excess evaluation algorithm. The algorithm is divided into the following steps, and a sketch map of the quantum circuit is shown in Fig.~\ref{fig:circuitqee}.

\textbf{Setup:}
As introduced in the main text, the algorithm aims to estimate $\sum_{X_j\geq 1} f(X_j) (X_j-M(X_j))$. We start this with representing $x$ with $n$ qubits $|x\rangle = |x_{n-1}\cdots x_1x_0\rangle$ under the condition that $x=\sum_{i=0}^{n-1}x_i2^i$, where $x_i\in\{0,1\},\forall i$. Then we re-scale the interval $[0,10]$ to $[0,2^n-1]$ such that the state $|x\rangle$ indeed represents the number $10x/(2^n-1)$ (For instance, $|0\rangle^{\otimes n}$ and $|1\rangle^{\otimes n}$ refer to 0 and 10, respectively).

\textbf{State preparation:} 
We use $n$ qubits to represent each loss $X_j$ as $|j\rangle$. Then, using quantum superposition, we prepare the quantum state
\begin{equation}
    |\psi_1\rangle = \sum_{j=0}^{N-1} \sqrt{p_j}\,|j\rangle
\end{equation}
with the unitary $U_X$, where $p_j=f(X_j)/C$ with $C$ a normalization factor.

\textbf{Quantum subtractor:} 
When $x\geq 1$, the payment is $x - M(x)=0.4(x-1)$. We first need to evaluate $x-1$ with a quantum Subtractor ($U_S$) detailed in \cite{Ruiz_Perez_2017}, in which the threshold 1 needs to be converted as $1_n=\text{bin}(\tfrac{2^n-1}{10})$. With one auxiliary qubit, we can finish this task and the resulting state is:
\begin{equation}
    |\psi_2\rangle = \sum_{j=0}^{1_n-1} \sqrt{p_j}\,|N-j-1_n\rangle |1\rangle_a + \sum_{j=1_n}^{N-1} \sqrt{p_j}\,|j-1_n\rangle |0\rangle_a,
\end{equation}
where $q_a$ means the auxiliary qubit.

\textbf{Controlled rotations:}
Apply an $X$ gate on the auxiliary qubit, and then add one new measurement qubit $q_m$, followed by an $RY(\pi/2)$ gate; we have:
\begin{equation}
\begin{aligned}
    |\psi_3\rangle =& \sum_{j=0}^{1_n-1} \sqrt{p_j}\,|N-x-1_n\rangle |0\rangle_a \left[\cos\!\bigl(\tfrac{\pi}{4}\bigr)|0\rangle_m + \sin\!\bigl(\tfrac{\pi}{4}\bigr)|1\rangle_m \right]\\
    &+ \sum_{j=1_n}^{N-1} \sqrt{p_j}\,|j-1_n\rangle |1\rangle_a \left[\cos\!\bigl(\tfrac{\pi}{4}\bigr)|0\rangle_m + \sin\!\bigl(\tfrac{\pi}{4}\bigr)|1\rangle_m\right].
\end{aligned}
\end{equation}
Consider the $CCRY(\theta)$ gate, which has two control qubits and one target qubit. It applies an $RY(\theta)$ gate on the target qubit only if the two control qubits are both in the state $|1\rangle$:
\begin{equation}
    CCRY(\theta)\,|1\rangle |1\rangle |\psi\rangle =   |1\rangle |1\rangle\,RY(\theta)\,|\psi\rangle, \qquad CCRY(\theta)\,|1\rangle |0\rangle |\psi\rangle =   |1\rangle |0\rangle |\psi\rangle.
\end{equation}
Then using $RY(0)=I$, we can combine the above two equations as
\begin{equation}\label{ccry}
    CCRY(\theta)\,|1\rangle |x\rangle |\psi\rangle =   |1\rangle |x\rangle\,RY(\theta x)\,|\psi\rangle.
\end{equation}
Then we apply a sequence of $n$ $CCRY$ gates shown in Fig.~\ref{fig:circuitqee}. For each gate, the first control qubit is the auxiliary qubit $q_a$. The second control qubit is one of the $n$ qubits that prepare $|j\rangle$. The rotation angle is $2^ic$, where $c$ is a relatively small positive number and $i$ is the index in the binary representation. According to Eq.~(\ref{ccry}), these operations will convert the state $|x_{n-1}\rangle \cdots |x_1\rangle |x_0\rangle |1\rangle |\psi\rangle$ to $|x_{n-1}\rangle \cdots |x_1\rangle |x_0\rangle |1\rangle 
RY\!\bigl(c\!\sum_{i=0}^{n-1} 2^i x_i \bigr)\,|\psi\rangle$, or more accurately, $|x_{n-1}\rangle \cdots |x_1\rangle |x_0\rangle |1\rangle 
RY(cx )\,|\psi\rangle$. Therefore, $|\psi_3\rangle$ is evolved to the following state:
\begin{equation}
\begin{aligned}
   |\psi_4\rangle =& \sum_{x=0}^{1_n-1}\sqrt{p_{x}}\;|N-x-1_{n}\rangle|0\rangle_a\bigl(\cos\tfrac{\pi}{4}|0\rangle_m+\sin\tfrac{\pi}{4}|1\rangle_m\bigr)\\
    &+\sum_{x=1_n}^{N-1}\sqrt{p_{x}}\;|x-1_{n}\rangle|1\rangle_a \otimes \bigl(\cos \theta_x |0\rangle_m+\sin \theta_x |1\rangle_m\bigr),
\end{aligned}
\end{equation}
where $\theta_x = \tfrac{\pi}{4}+c(x-1_n)$.

\textbf{Measure and post-processing:} Finally, we measure the measurement qubit $q_m$, and the probability of obtaining $|1\rangle$ is
\begin{equation}
    P_0 = \sum_{x=0}^{1_n-1} p_{x}\,\sin^{2}\!\bigl(\tfrac{\pi}{4}\bigr)+\sum_{x=1_n}^{N-1} p_{x}\,\sin^{2}\!\bigl(\tfrac{\pi}{4}+c(x-1_n)\bigr).
\end{equation}
When $c$ is very small, we use Taylor's expansion and neglect terms of order $\mathcal{O}(c^2)$; the result is $\sin^{2}\!\bigl(\tfrac{\pi}{4}+c(x-1_n)\bigr) \approx c(x-1_n)+\tfrac12$. Finally, the above probability is:
\begin{equation}
    P_0 = \sum_{x=0}^{1_n-1} \tfrac12 p_{x}+\sum_{x=1_n}^{N-1}p_{x}\Bigl(\tfrac12 + c(x-1_n)\Bigr) = \tfrac12 \sum_{x=0}^{N-1} p_x +c \sum_{x=1_n}^{N-1} p_x (x-1_n),
\end{equation}
With $\sum_x p_x=1$, it follows that:
\begin{equation}
     \sum_{x=1_n}^{N-1} p_x (x-1_n) = \frac{P_0 - \tfrac12}{c}.
\end{equation}
We can see that the left-hand side is exactly what we need:
\begin{equation}
    \sum_{X_j\geq 1} f(X_j) \bigl(X_j - M(X_j)\bigr) = 0.4 \times \frac{10}{2^n-1}  \sum_{x=1_n}^{N-1} p_x (x-1_n).
\end{equation}
Then the algorithm is finished.

\section{Quantum singular value decomposition algorithm}
\label{method:qsvd}

In this part, we briefly introduce the QSVD algorithm, which is detailed in \cite{Bravo2020quantum}.

\textbf{Vectorized SVD:}
The SVD of a matrix \(D\) is
\begin{equation}
    D=\sum_i \sigma_i |u_i\rangle \langle v_i|,
\end{equation}
where \(\sigma_i\) is the singular value, and \(|u_i\rangle\) or \(|v_i\rangle\) is the associated left or right singular vector. We can also express it using a vectorized matrix representation, which employs a vector of \(MN\) elements to represent the \(M \times N\) matrix:
\begin{equation}
    D = \begin{pmatrix}
        D_{11} & \cdots & D_{1N} \\
        \vdots &       & \vdots  \\
        D_{M1} & \cdots & D_{MM} 
    \end{pmatrix}  
    \; \to \;
    |D\rangle = \begin{pmatrix}
        D_{11} & \cdots & D_{1N} & \cdots & D_{M1} & \cdots & D_{MM}
    \end{pmatrix}^{\mathrm{T}}.
\end{equation}
Therefore, the SVD of a vectorized matrix is expressed as
\begin{equation}
    D =\sum_i \sigma_i |u_i\rangle \langle v_i|
    \;\;\to\;\;
    |D\rangle = \sum_i \sigma_i |u_i\rangle |v_i\rangle.
\end{equation}

\textbf{QSVD algorithm:}
In quantum computers, with a re-normalization factor \(C = \sqrt{\langle D|D\rangle}\), we can prepare the quantum state of \(D\) and express it in the Schmidt decomposition form
\begin{equation}
    |\psi\rangle_{AB} = \frac{1}{C} |D\rangle 
    = \sum_i \lambda_i \,|u_i\rangle_A \,|v_i\rangle_B,
\end{equation}
where \(A,B\) represent the two subsystems. Consider two unitary operations \(U\) and \(V\) that convert the states \(\{|u_i\rangle\}\) and \(\{|v_i\rangle\}\) to the computational basis \(\{|e_i\rangle\}\):
\begin{equation}
    U|u_i\rangle = e^{ip_i} |e_i\rangle, 
    \qquad 
    V|v_i\rangle = e^{iq_i} |e_i\rangle.
\end{equation}
We have
\begin{equation}
    (U_A\otimes V_B)|\psi\rangle_{AB} 
    = \sum_i \lambda_i \, e^{i(p_i+q_i)} \,|e_i\rangle_A \,|e_i\rangle_B.
\end{equation}
Therefore, when measuring the two subsystems \(A\) and \(B\) of the state \(|\psi\rangle_{AB}\), we obtain the same measurement result on the computational basis. The corresponding probability is \(\lambda_i^2 = \sigma_i^2 / C^2\). Finally, applying \(U^\dagger\) and \(V^\dagger\) on the computational basis state \(|e_i\rangle\) yields the singular vectors \(|u_i\rangle\) and \(|v_i\rangle\).

To obtain \(U\) and \(V\), we apply the variational QSVD algorithm by introducing an optimization procedure. That is, after preparing the state \(|\psi\rangle_{AB}\), we apply two parameterized quantum circuits \(U_A(\bm{\theta}_1)\) and \(V_B(\bm{\theta}_2)\). The loss function is
\begin{equation}
    C = \frac{1}{N_m} \sum_{x=1}^{N_m} \delta(p_x, q_x),
\end{equation}
where \(N_m\) is the number of shots, and \(p_x, q_x\) are the two binary strings at the \(x\)-th shot.
\(\delta(a,b)\) is 0 if \(a \neq b\) and 1 if \(a = b\). A classical optimizer is employed to optimize the parameters to minimize the loss function.

\end{appendices}

\bibliography{ref}


\begin{thebibliography}{97}
\ifx \bisbn   \undefined \def \bisbn  #1{ISBN #1}\fi
\ifx \binits  \undefined \def \binits#1{#1}\fi
\ifx \bauthor  \undefined \def \bauthor#1{#1}\fi
\ifx \batitle  \undefined \def \batitle#1{#1}\fi
\ifx \bjtitle  \undefined \def \bjtitle#1{#1}\fi
\ifx \bvolume  \undefined \def \bvolume#1{\textbf{#1}}\fi
\ifx \byear  \undefined \def \byear#1{#1}\fi
\ifx \bissue  \undefined \def \bissue#1{#1}\fi
\ifx \bfpage  \undefined \def \bfpage#1{#1}\fi
\ifx \blpage  \undefined \def \blpage #1{#1}\fi
\ifx \burl  \undefined \def \burl#1{\textsf{#1}}\fi
\ifx \doiurl  \undefined \def \doiurl#1{\url{https://doi.org/#1}}\fi
\ifx \betal  \undefined \def \betal{\textit{et al.}}\fi
\ifx \binstitute  \undefined \def \binstitute#1{#1}\fi
\ifx \binstitutionaled  \undefined \def \binstitutionaled#1{#1}\fi
\ifx \bctitle  \undefined \def \bctitle#1{#1}\fi
\ifx \beditor  \undefined \def \beditor#1{#1}\fi
\ifx \bpublisher  \undefined \def \bpublisher#1{#1}\fi
\ifx \bbtitle  \undefined \def \bbtitle#1{#1}\fi
\ifx \bedition  \undefined \def \bedition#1{#1}\fi
\ifx \bseriesno  \undefined \def \bseriesno#1{#1}\fi
\ifx \blocation  \undefined \def \blocation#1{#1}\fi
\ifx \bsertitle  \undefined \def \bsertitle#1{#1}\fi
\ifx \bsnm \undefined \def \bsnm#1{#1}\fi
\ifx \bsuffix \undefined \def \bsuffix#1{#1}\fi
\ifx \bparticle \undefined \def \bparticle#1{#1}\fi
\ifx \barticle \undefined \def \barticle#1{#1}\fi
\bibcommenthead
\ifx \bconfdate \undefined \def \bconfdate #1{#1}\fi
\ifx \botherref \undefined \def \botherref #1{#1}\fi
\ifx \url \undefined \def \url#1{\textsf{#1}}\fi
\ifx \bchapter \undefined \def \bchapter#1{#1}\fi
\ifx \bbook \undefined \def \bbook#1{#1}\fi
\ifx \bcomment \undefined \def \bcomment#1{#1}\fi
\ifx \oauthor \undefined \def \oauthor#1{#1}\fi
\ifx \citeauthoryear \undefined \def \citeauthoryear#1{#1}\fi
\ifx \endbibitem  \undefined \def \endbibitem {}\fi
\ifx \bconflocation  \undefined \def \bconflocation#1{#1}\fi
\ifx \arxivurl  \undefined \def \arxivurl#1{\textsf{#1}}\fi
\csname PreBibitemsHook\endcsname

\bibitem[\protect\citeauthoryear{Ewold}{1991}]{ewold1991insurance}
\begin{barticle}
\bauthor{\bsnm{Ewold}, \binits{F.}}:
\batitle{Insurance and risk}.
\bjtitle{The Foucault effect: Studies in governmentality}
\bvolume{197210},
\bfpage{201}--\blpage{202}
(\byear{1991})
\end{barticle}
\endbibitem

\bibitem[\protect\citeauthoryear{Vaughan and
  Vaughan}{2007}]{vaughan2007fundamentals}
\begin{bbook}
\bauthor{\bsnm{Vaughan}, \binits{E.J.}},
\bauthor{\bsnm{Vaughan}, \binits{T.}}:
\bbtitle{Fundamentals of Risk and Insurance}.
\bpublisher{John Wiley \& Sons},
\blocation{Hoboken}
(\byear{2007})
\end{bbook}
\endbibitem

\bibitem[\protect\citeauthoryear{Dickson}{2016}]{dickson2016insurance}
\begin{bbook}
\bauthor{\bsnm{Dickson}, \binits{D.C.M.}}:
\bbtitle{Insurance Risk and Ruin}.
\bpublisher{Cambridge University Press},
\blocation{Cambridge}
(\byear{2016}).
\burl{https://books.google.com/books?id=Ku08DQAAQBAJ}
\end{bbook}
\endbibitem

\bibitem[\protect\citeauthoryear{Frees}{2009}]{frees2009regression}
\begin{bbook}
\bauthor{\bsnm{Frees}, \binits{E.W.}}:
\bbtitle{Regression Modeling with Actuarial and Financial Applications}.
\bpublisher{Cambridge University Press},
\blocation{Cambridge}
(\byear{2009})
\end{bbook}
\endbibitem

\bibitem[\protect\citeauthoryear{Ohlsson and Johansson}{2010}]{ohlsson2010non}
\begin{bbook}
\bauthor{\bsnm{Ohlsson}, \binits{E.}},
\bauthor{\bsnm{Johansson}, \binits{B.}}:
\bbtitle{Non-life Insurance Pricing with Generalized Linear Models}
vol. \bseriesno{2}.
\bpublisher{Springer},
\blocation{Heidelberg}
(\byear{2010})
\end{bbook}
\endbibitem

\bibitem[\protect\citeauthoryear{Richman}{2018}]{richman2018ai}
\begin{botherref}
\oauthor{\bsnm{Richman}, \binits{R.}}:
Ai in actuarial science.
Available at SSRN 3218082
(2018)
\end{botherref}
\endbibitem

\bibitem[\protect\citeauthoryear{Promislow}{2014}]{promislow2014fundamentals}
\begin{bbook}
\bauthor{\bsnm{Promislow}, \binits{S.D.}}:
\bbtitle{Fundamentals of Actuarial Mathematics}.
\bpublisher{John Wiley \& Sons},
\blocation{Hoboken}
(\byear{2014})
\end{bbook}
\endbibitem

\bibitem[\protect\citeauthoryear{Lee}{2000}]{lee2000lee}
\begin{barticle}
\bauthor{\bsnm{Lee}, \binits{R.}}:
\batitle{The lee-carter method for forecasting mortality, with various
  extensions and applications}.
\bjtitle{North American actuarial journal}
\bvolume{4}(\bissue{1}),
\bfpage{80}--\blpage{91}
(\byear{2000})
\end{barticle}
\endbibitem

\bibitem[\protect\citeauthoryear{Cairns et~al.}{2006}]{cairns2006two}
\begin{barticle}
\bauthor{\bsnm{Cairns}, \binits{A.J.G.}},
\bauthor{\bsnm{Blake}, \binits{D.}},
\bauthor{\bsnm{Dowd}, \binits{K.}}:
\batitle{A two-factor model for stochastic mortality with parameter
  uncertainty: Theory and calibration}.
\bjtitle{Journal of Risk and Insurance}
\bvolume{73}(\bissue{4}),
\bfpage{687}--\blpage{718}
(\byear{2006})
\doiurl{10.1111/j.1539-6975.2006.00195.x}
\end{barticle}
\endbibitem

\bibitem[\protect\citeauthoryear{Brown et~al.}{2001}]{brown2001role}
\begin{bbook}
\bauthor{\bsnm{Brown}, \binits{J.R.}},
\bauthor{\bsnm{Mitchell}, \binits{O.S.}},
\bauthor{\bsnm{Poterba}, \binits{J.M.}},
\bauthor{\bsnm{Warshawsky}, \binits{M.J.}}:
\bbtitle{The Role of Annuity Markets in Financing Retirement}.
\bpublisher{Mit Press},
\blocation{Massachusetts}
(\byear{2001})
\end{bbook}
\endbibitem

\bibitem[\protect\citeauthoryear{Embrechts and
  W{\"u}thrich}{2022}]{embrechts2022recent}
\begin{barticle}
\bauthor{\bsnm{Embrechts}, \binits{P.}},
\bauthor{\bsnm{W{\"u}thrich}, \binits{M.V.}}:
\batitle{Recent challenges in actuarial science}.
\bjtitle{Annual Review of Statistics and Its Application}
\bvolume{9}(\bissue{1}),
\bfpage{1}--\blpage{22}
(\byear{2022})
\end{barticle}
\endbibitem

\bibitem[\protect\citeauthoryear{Delong et~al.}{2019a}]{delong2019fair1}
\begin{barticle}
\bauthor{\bsnm{Delong}, \binits{{\L}.}},
\bauthor{\bsnm{Dhaene}, \binits{J.}},
\bauthor{\bsnm{Barigou}, \binits{K.}}:
\batitle{Fair valuation of insurance liability cash-flow streams in continuous
  time: Applications}.
\bjtitle{ASTIN Bulletin: The Journal of the IAA}
\bvolume{49}(\bissue{2}),
\bfpage{299}--\blpage{333}
(\byear{2019})
\end{barticle}
\endbibitem

\bibitem[\protect\citeauthoryear{Delong et~al.}{2019b}]{delong2019fair2}
\begin{barticle}
\bauthor{\bsnm{Delong}, \binits{{\L}.}},
\bauthor{\bsnm{Dhaene}, \binits{J.}},
\bauthor{\bsnm{Barigou}, \binits{K.}}:
\batitle{Fair valuation of insurance liability cash-flow streams in continuous
  time: Theory}.
\bjtitle{Insurance: Mathematics and Economics}
\bvolume{88},
\bfpage{196}--\blpage{208}
(\byear{2019})
\end{barticle}
\endbibitem

\bibitem[\protect\citeauthoryear{Albrecher
  et~al.}{2017}]{albrecher2017reinsurance}
\begin{bbook}
\bauthor{\bsnm{Albrecher}, \binits{H.}},
\bauthor{\bsnm{Beirlant}, \binits{J.}},
\bauthor{\bsnm{Teugels}, \binits{J.L.}}:
\bbtitle{Reinsurance: Actuarial and Statistical Aspects}.
\bpublisher{John Wiley \& Sons},
\blocation{Hoboken}
(\byear{2017})
\end{bbook}
\endbibitem

\bibitem[\protect\citeauthoryear{Ladoucette and
  Teugels}{2006}]{ladoucette2006reinsurance}
\begin{barticle}
\bauthor{\bsnm{Ladoucette}, \binits{S.A.}},
\bauthor{\bsnm{Teugels}, \binits{J.L.}}:
\batitle{Reinsurance of large claims}.
\bjtitle{Journal of Computational and Applied Mathematics}
\bvolume{186}(\bissue{1}),
\bfpage{163}--\blpage{190}
(\byear{2006})
\end{barticle}
\endbibitem

\bibitem[\protect\citeauthoryear{Dickson and
  Waters}{1996}]{dickson1996reinsurance}
\begin{barticle}
\bauthor{\bsnm{Dickson}, \binits{D.C.}},
\bauthor{\bsnm{Waters}, \binits{H.R.}}:
\batitle{Reinsurance and ruin}.
\bjtitle{Insurance: Mathematics and Economics}
\bvolume{19}(\bissue{1}),
\bfpage{61}--\blpage{80}
(\byear{1996})
\end{barticle}
\endbibitem

\bibitem[\protect\citeauthoryear{Hey}{2023}]{Hey2023Book}
\begin{bbook}
\bauthor{\bsnm{Hey}, \binits{T.}}:
\bbtitle{Feynman Lectures on Computation: Anniversary Edition (2nd Ed.).}
\bpublisher{CRS Press},
\blocation{Boca Raton}
(\byear{2023}).
\doiurl{10.1201/9781003358817}
\end{bbook}
\endbibitem

\bibitem[\protect\citeauthoryear{McArdle et~al.}{2020}]{mcardle2020quantum}
\begin{barticle}
\bauthor{\bsnm{McArdle}, \binits{S.}},
\bauthor{\bsnm{Endo}, \binits{S.}},
\bauthor{\bsnm{Aspuru-Guzik}, \binits{A.}},
\bauthor{\bsnm{Benjamin}, \binits{S.C.}},
\bauthor{\bsnm{Yuan}, \binits{X.}}:
\batitle{Quantum computational chemistry}.
\bjtitle{Rev. Mod. Phys.}
\bvolume{92},
\bfpage{015003}
(\byear{2020})
\doiurl{10.1103/RevModPhys.92.015003}
\end{barticle}
\endbibitem

\bibitem[\protect\citeauthoryear{Outeiral et~al.}{2021}]{outeiral2021prospects}
\begin{barticle}
\bauthor{\bsnm{Outeiral}, \binits{C.}},
\bauthor{\bsnm{Strahm}, \binits{M.}},
\bauthor{\bsnm{Shi}, \binits{J.}},
\bauthor{\bsnm{Morris}, \binits{G.M.}},
\bauthor{\bsnm{Benjamin}, \binits{S.C.}},
\bauthor{\bsnm{Deane}, \binits{C.M.}}:
\batitle{The prospects of quantum computing in computational molecular
  biology}.
\bjtitle{WIREs Computational Molecular Science}
\bvolume{11}(\bissue{1}),
\bfpage{1481}
(\byear{2021})
\doiurl{10.1002/wcms.1481}
\end{barticle}
\endbibitem

\bibitem[\protect\citeauthoryear{Emani et~al.}{2021}]{emani2021quantum}
\begin{barticle}
\bauthor{\bsnm{Emani}, \binits{P.S.}},
\bauthor{\bsnm{Warrell}, \binits{J.}},
\bauthor{\bsnm{Anticevic}, \binits{A.}},
\bauthor{\bsnm{Bekiranov}, \binits{S.}},
\bauthor{\bsnm{Gandal}, \binits{M.}},
\bauthor{\bsnm{McConnell}, \binits{M.J.}},
\bauthor{\bsnm{Sapiro}, \binits{G.}},
\bauthor{\bsnm{Aspuru-Guzik}, \binits{A.}},
\bauthor{\bsnm{Baker}, \binits{J.T.}},
\bauthor{\bsnm{Bastiani}, \binits{M.}},
\bauthor{\bsnm{Murray}, \binits{J.D.}},
\bauthor{\bsnm{Sotiropoulos}, \binits{S.N.}},
\bauthor{\bsnm{Taylor}, \binits{J.}},
\bauthor{\bsnm{Senthil}, \binits{G.}},
\bauthor{\bsnm{Lehner}, \binits{T.}},
\bauthor{\bsnm{Gerstein}, \binits{M.B.}},
\bauthor{\bsnm{Harrow}, \binits{A.W.}}:
\batitle{Quantum computing at the frontiers of biological sciences}.
\bjtitle{Nature Methods}
\bvolume{18}(\bissue{7}),
\bfpage{701}--\blpage{709}
(\byear{2021})
\doiurl{10.1038/s41592-020-01004-3}
\end{barticle}
\endbibitem

\bibitem[\protect\citeauthoryear{Ma et~al.}{2020}]{ma2020quantum}
\begin{barticle}
\bauthor{\bsnm{Ma}, \binits{H.}},
\bauthor{\bsnm{Govoni}, \binits{M.}},
\bauthor{\bsnm{Galli}, \binits{G.}}:
\batitle{Quantum simulations of materials on near-term quantum computers}.
\bjtitle{npj Computational Materials}
\bvolume{6}(\bissue{1}),
\bfpage{85}
(\byear{2020})
\doiurl{10.1038/s41524-020-00353-z}
\end{barticle}
\endbibitem

\bibitem[\protect\citeauthoryear{Cao et~al.}{2018}]{cao2018potential}
\begin{barticle}
\bauthor{\bsnm{Cao}, \binits{Y.}},
\bauthor{\bsnm{Romero}, \binits{J.}},
\bauthor{\bsnm{Aspuru-Guzik}, \binits{A.}}:
\batitle{Potential of quantum computing for drug discovery}.
\bjtitle{IBM Journal of Research and Development}
\bvolume{62}(\bissue{6}),
\bfpage{6}--\blpage{1620}
(\byear{2018})
\doiurl{10.1147/JRD.2018.2888987}
\end{barticle}
\endbibitem

\bibitem[\protect\citeauthoryear{Orús et~al.}{2019}]{orus2019quantum}
\begin{barticle}
\bauthor{\bsnm{Orús}, \binits{R.}},
\bauthor{\bsnm{Mugel}, \binits{S.}},
\bauthor{\bsnm{Lizaso}, \binits{E.}}:
\batitle{Quantum computing for finance: Overview and prospects}.
\bjtitle{Reviews in Physics}
\bvolume{4},
\bfpage{100028}
(\byear{2019})
\doiurl{10.1016/j.revip.2019.100028}
\end{barticle}
\endbibitem

\bibitem[\protect\citeauthoryear{Egger et~al.}{2020}]{egger2020quantum}
\begin{barticle}
\bauthor{\bsnm{Egger}, \binits{D.J.}},
\bauthor{\bsnm{Gambella}, \binits{C.}},
\bauthor{\bsnm{Marecek}, \binits{J.}},
\bauthor{\bsnm{McFaddin}, \binits{S.}},
\bauthor{\bsnm{Mevissen}, \binits{M.}},
\bauthor{\bsnm{Raymond}, \binits{R.}},
\bauthor{\bsnm{Simonetto}, \binits{A.}},
\bauthor{\bsnm{Woerner}, \binits{S.}},
\bauthor{\bsnm{Yndurain}, \binits{E.}}:
\batitle{Quantum computing for finance: State-of-the-art and future prospects}.
\bjtitle{IEEE Transactions on Quantum Engineering}
\bvolume{1},
\bfpage{1}--\blpage{24}
(\byear{2020})
\doiurl{10.1109/TQE.2020.3030314}
\end{barticle}
\endbibitem

\bibitem[\protect\citeauthoryear{Herman et~al.}{2022}]{herman2022survey}
\begin{botherref}
\oauthor{\bsnm{Herman}, \binits{D.}},
\oauthor{\bsnm{Googin}, \binits{C.}},
\oauthor{\bsnm{Liu}, \binits{X.}},
\oauthor{\bsnm{Galda}, \binits{A.}},
\oauthor{\bsnm{Safro}, \binits{I.}},
\oauthor{\bsnm{Sun}, \binits{Y.}},
\oauthor{\bsnm{Pistoia}, \binits{M.}},
\oauthor{\bsnm{Alexeev}, \binits{Y.}}:
A survey of quantum computing for finance
(2022)
{\href{https://arxiv.org/abs/2201.02773}{{arXiv:2201.02773}}}
{[quant-ph]}
\end{botherref}
\endbibitem

\bibitem[\protect\citeauthoryear{Adam}{2022}]{adam2022potential}
\begin{botherref}
\oauthor{\bsnm{Adam}, \binits{M.}}:
Potential applications of quantum computing for the insurance industry
(2022)
{\href{https://arxiv.org/abs/2210.06172}{{arXiv:2210.06172}}}
{[quant-ph]}
\end{botherref}
\endbibitem

\bibitem[\protect\citeauthoryear{Zhuang et~al.}{2023}]{zhuang2022quantum1}
\begin{barticle}
\bauthor{\bsnm{Zhuang}, \binits{X.-N.}},
\bauthor{\bsnm{Chen}, \binits{Z.-Y.}},
\bauthor{\bsnm{Xue}, \binits{C.}},
\bauthor{\bsnm{Wu}, \binits{Y.-C.}},
\bauthor{\bsnm{Guo}, \binits{G.-P.}}:
\batitle{Quantum {E}ncoding and {A}nalysis on {C}ontinuous {T}ime {S}tochastic
  {P}rocess with {F}inancial {A}pplications}.
\bjtitle{{Quantum}}
\bvolume{7},
\bfpage{1127}
(\byear{2023})
\doiurl{10.22331/q-2023-10-03-1127}
\end{barticle}
\endbibitem

\bibitem[\protect\citeauthoryear{Tamturk and Carenzo}{2023}]{capital1}
\begin{barticle}
\bauthor{\bsnm{Tamturk}, \binits{M.}},
\bauthor{\bsnm{Carenzo}, \binits{M.}}:
\batitle{Quantum computing in insurance capital modelling under reinsurance
  contracts}.
\bjtitle{Appliedmath}
\bvolume{3}(\bissue{4}),
\bfpage{741}--\blpage{757}
(\byear{2023})
\doiurl{10.3390/appliedmath3040040}
\end{barticle}
\endbibitem

\bibitem[\protect\citeauthoryear{Tamturk}{2023}]{capital2}
\begin{botherref}
\oauthor{\bsnm{Tamturk}, \binits{M.}}:
Quantum computing in insurance capital modelling.
Mathematics
\textbf{11}(3)
(2023)
\doiurl{10.3390/math11030658}
\end{botherref}
\endbibitem

\bibitem[\protect\citeauthoryear{Naik and Bhise}{}]{riskide}
\begin{botherref}
\oauthor{\bsnm{Naik}, \binits{K.S.}},
\oauthor{\bsnm{Bhise}, \binits{A.}}:
Risk identification using quantum machine learning for fleet insurance premium.
In: Panda, S.K., Rout, R.R., Sadam, R.C., Rayanoothala, B.V.S., Li, K.-C.,
  Buyya, R. (eds.)
Computing, Communication and Learning,
pp. 277--288.
Springer
\end{botherref}
\endbibitem

\bibitem[\protect\citeauthoryear{Lefevre et~al.}{2024}]{cber}
\begin{botherref}
\oauthor{\bsnm{Lefevre}, \binits{C.}},
\oauthor{\bsnm{Tamturk}, \binits{M.}},
\oauthor{\bsnm{Utev}, \binits{S.}},
\oauthor{\bsnm{Carenzo}, \binits{M.}}:
Cyber risk in insurance: A quantum modeling.
Risks
\textbf{12}(5)
(2024)
\doiurl{10.3390/risks12050083}
\end{botherref}
\endbibitem

\bibitem[\protect\citeauthoryear{Pushpak and Jain}{2022}]{fraud}
\begin{bchapter}
\bauthor{\bsnm{Pushpak}, \binits{S.N.}},
\bauthor{\bsnm{Jain}, \binits{S.}}:
\bctitle{An implementation of quantum machine learning technique to determine
  insurance claim fraud}.
In: \bbtitle{2022 10th International Conference on Reliability, Infocom
  Technologies and Optimization (Trends and Future Directions) (ICRITO)},
pp. \bfpage{1}--\blpage{5}
(\byear{2022}).
\doiurl{10.1109/ICRITO56286.2022.9964828}
\end{bchapter}
\endbibitem

\bibitem[\protect\citeauthoryear{Beard}{2013}]{beard2013risk}
\begin{bbook}
\bauthor{\bsnm{Beard}, \binits{R.}}:
\bbtitle{Risk Theory: the Stochastic Basis of Insurance}
vol. \bseriesno{20}.
\bpublisher{Springer},
\blocation{Heidelberg}
(\byear{2013})
\end{bbook}
\endbibitem

\bibitem[\protect\citeauthoryear{B{\"u}hlmann}{2007}]{buhlmann2007mathematical}
\begin{bbook}
\bauthor{\bsnm{B{\"u}hlmann}, \binits{H.}}:
\bbtitle{Mathematical Methods in Risk Theory}
vol. \bseriesno{172}.
\bpublisher{Springer},
\blocation{Heidelberg}
(\byear{2007})
\end{bbook}
\endbibitem

\bibitem[\protect\citeauthoryear{Mayer and Butler}{1993}]{mayer1993statistical}
\begin{barticle}
\bauthor{\bsnm{Mayer}, \binits{D.}},
\bauthor{\bsnm{Butler}, \binits{D.}}:
\batitle{Statistical validation}.
\bjtitle{Ecological modelling}
\bvolume{68}(\bissue{1-2}),
\bfpage{21}--\blpage{32}
(\byear{1993})
\end{barticle}
\endbibitem

\bibitem[\protect\citeauthoryear{W{\"u}thrich and
  Merz}{2013}]{wuthrich2013financial}
\begin{botherref}
\oauthor{\bsnm{W{\"u}thrich}, \binits{M.V.}},
\oauthor{\bsnm{Merz}, \binits{M.}}:
Financial modeling, actuarial valuation and solvency in insurance.
Technical report,
Springer
(2013)
\end{botherref}
\endbibitem

\bibitem[\protect\citeauthoryear{Gupta et~al.}{2006}]{gupta2006model}
\begin{botherref}
\oauthor{\bsnm{Gupta}, \binits{H.V.}},
\oauthor{\bsnm{Beven}, \binits{K.J.}},
\oauthor{\bsnm{Wagener}, \binits{T.}}:
Model calibration and uncertainty estimation.
Encyclopedia of hydrological sciences
(2006)
\end{botherref}
\endbibitem

\bibitem[\protect\citeauthoryear{C{\'o}rcoles
  et~al.}{2019}]{corcoles2019challenges}
\begin{barticle}
\bauthor{\bsnm{C{\'o}rcoles}, \binits{A.D.}},
\bauthor{\bsnm{Kandala}, \binits{A.}},
\bauthor{\bsnm{Javadi-Abhari}, \binits{A.}},
\bauthor{\bsnm{McClure}, \binits{D.T.}},
\bauthor{\bsnm{Cross}, \binits{A.W.}},
\bauthor{\bsnm{Temme}, \binits{K.}},
\bauthor{\bsnm{Nation}, \binits{P.D.}},
\bauthor{\bsnm{Steffen}, \binits{M.}},
\bauthor{\bsnm{Gambetta}, \binits{J.M.}}:
\batitle{Challenges and opportunities of near-term quantum computing systems}.
\bjtitle{Proceedings of the IEEE}
\bvolume{108}(\bissue{8}),
\bfpage{1338}--\blpage{1352}
(\byear{2019})
\end{barticle}
\endbibitem

\bibitem[\protect\citeauthoryear{Weigold et~al.}{2020}]{weigold2020data}
\begin{bchapter}
\bauthor{\bsnm{Weigold}, \binits{M.}},
\bauthor{\bsnm{Barzen}, \binits{J.}},
\bauthor{\bsnm{Leymann}, \binits{F.}},
\bauthor{\bsnm{Salm}, \binits{M.}}:
\bctitle{Data encoding patterns for quantum computing}.
In: \bbtitle{Proceedings of the 27th Conference on Pattern Languages of
  Programs},
pp. \bfpage{1}--\blpage{11}
(\byear{2020})
\end{bchapter}
\endbibitem

\bibitem[\protect\citeauthoryear{Zhang et~al.}{2022}]{zhang2022quantum}
\begin{barticle}
\bauthor{\bsnm{Zhang}, \binits{X.-M.}},
\bauthor{\bsnm{Li}, \binits{T.}},
\bauthor{\bsnm{Yuan}, \binits{X.}}:
\batitle{Quantum state preparation with optimal circuit depth: Implementations
  and applications}.
\bjtitle{Physical Review Letters}
\bvolume{129}(\bissue{23}),
\bfpage{230504}
(\byear{2022})
\end{barticle}
\endbibitem

\bibitem[\protect\citeauthoryear{Lau et~al.}{2022}]{lau2022nisq}
\begin{barticle}
\bauthor{\bsnm{Lau}, \binits{J.W.Z.}},
\bauthor{\bsnm{Lim}, \binits{K.H.}},
\bauthor{\bsnm{Shrotriya}, \binits{H.}},
\bauthor{\bsnm{Kwek}, \binits{L.C.}}:
\batitle{Nisq computing: where are we and where do we go?}
\bjtitle{AAPPS bulletin}
\bvolume{32}(\bissue{1}),
\bfpage{27}
(\byear{2022})
\end{barticle}
\endbibitem

\bibitem[\protect\citeauthoryear{Preskill}{2018}]{preskill2018quantum}
\begin{barticle}
\bauthor{\bsnm{Preskill}, \binits{J.}}:
\batitle{Quantum {C}omputing in the {NISQ} era and beyond}.
\bjtitle{{Quantum}}
\bvolume{2},
\bfpage{79}
(\byear{2018})
\doiurl{10.22331/q-2018-08-06-79}
\end{barticle}
\endbibitem

\bibitem[\protect\citeauthoryear{Campbell et~al.}{2017}]{campbell2017roads}
\begin{barticle}
\bauthor{\bsnm{Campbell}, \binits{E.T.}},
\bauthor{\bsnm{Terhal}, \binits{B.M.}},
\bauthor{\bsnm{Vuillot}, \binits{C.}}:
\batitle{Roads towards fault-tolerant universal quantum computation}.
\bjtitle{Nature}
\bvolume{549}(\bissue{7671}),
\bfpage{172}--\blpage{179}
(\byear{2017})
\doiurl{10.1038/nature23460}
\end{barticle}
\endbibitem

\bibitem[\protect\citeauthoryear{}{}]{wukong}
\begin{botherref}
Https://originqc.com.cn/.
\url{https://originqc.com.cn/}
\end{botherref}
\endbibitem

\bibitem[\protect\citeauthoryear{Cossette
  et~al.}{2019}]{Cossette2019Collective}
\begin{barticle}
\bauthor{\bsnm{Cossette}, \binits{H.}},
\bauthor{\bsnm{Marceau}, \binits{E.}},
\bauthor{\bsnm{Mtalai}, \binits{I.}}:
\batitle{Collective risk models with dependence}.
\bjtitle{Insurance: Mathematics and Economics}
\bvolume{87},
\bfpage{153}--\blpage{168}
(\byear{2019})
\end{barticle}
\endbibitem

\bibitem[\protect\citeauthoryear{de~Lourdes~Centeno}{2005}]{Lourdes2005Dependent}
\begin{barticle}
\bauthor{\bsnm{Lourdes~Centeno}, \binits{M.}}:
\batitle{Dependent risks and excess of loss reinsurance}.
\bjtitle{Insurance: Mathematics and Economics}
\bvolume{37}(\bissue{2}),
\bfpage{229}--\blpage{238}
(\byear{2005})
\end{barticle}
\endbibitem

\bibitem[\protect\citeauthoryear{Fahrmeir}{1992}]{Fahrmeir1992Posterior}
\begin{barticle}
\bauthor{\bsnm{Fahrmeir}, \binits{L.}}:
\batitle{Posterior mode estimation by extended kalman filtering for
  multivariate dynamic generalized linear models}.
\bjtitle{Journal of the American Statistical Association}
\bvolume{87}(\bissue{418}),
\bfpage{501}--\blpage{509}
(\byear{1992})
\doiurl{10.1080/01621459.1992.10475232} .
\bcomment{doi: 10.1080/01621459.1992.10475232}
\end{barticle}
\endbibitem

\bibitem[\protect\citeauthoryear{Aiuppa and
  Trieschmann}{1987}]{Aiuppa1987Empirical}
\begin{barticle}
\bauthor{\bsnm{Aiuppa}, \binits{T.A.}},
\bauthor{\bsnm{Trieschmann}, \binits{J.S.}}:
\batitle{An empirical analysis of the magnitude and accuracy of
  incurred-but-not-reported reserves}.
\bjtitle{The Journal of Risk and Insurance}
\bvolume{54}(\bissue{1}),
\bfpage{100}--\blpage{118}
(\byear{1987})
\doiurl{10.2307/252884}
\end{barticle}
\endbibitem

\bibitem[\protect\citeauthoryear{Renshaw and Verrall}{1998}]{Renshaw1998CLM}
\begin{barticle}
\bauthor{\bsnm{Renshaw}, \binits{A.E.}},
\bauthor{\bsnm{Verrall}, \binits{R.J.}}:
\batitle{A stochastic model underlying the chain-ladder technique}.
\bjtitle{British Actuarial Journal}
\bvolume{4}(\bissue{4}),
\bfpage{903}--\blpage{923}
(\byear{1998})
\doiurl{10.1017/S1357321700000222}
\end{barticle}
\endbibitem

\bibitem[\protect\citeauthoryear{Grossi et~al.}{2005}]{grossi2005catastrophe}
\begin{bbook}
\bauthor{\bsnm{Grossi}, \binits{P.}},
\bauthor{\bsnm{Kunreuther}, \binits{H.}},
\bauthor{\bsnm{Patel}, \binits{C.C.}}:
\bbtitle{Catastrophe Modeling: a New Approach to Managing Risk}
vol. \bseriesno{25}.
\bpublisher{Springer},
\blocation{Berlin}
(\byear{2005})
\end{bbook}
\endbibitem

\bibitem[\protect\citeauthoryear{Dvorkin et~al.}{2019}]{8515058}
\begin{barticle}
\bauthor{\bsnm{Dvorkin}, \binits{V.}},
\bauthor{\bsnm{Delikaraoglou}, \binits{S.}},
\bauthor{\bsnm{Morales}, \binits{J.M.}}:
\batitle{Setting reserve requirements to approximate the efficiency of the
  stochastic dispatch}.
\bjtitle{IEEE Transactions on Power Systems}
\bvolume{34}(\bissue{2}),
\bfpage{1524}--\blpage{1536}
(\byear{2019})
\doiurl{10.1109/TPWRS.2018.2878723}
\end{barticle}
\endbibitem

\bibitem[\protect\citeauthoryear{Stewart}{1971}]{Stewart1971Assessment}
\begin{barticle}
\bauthor{\bsnm{Stewart}, \binits{C.M.}}:
\batitle{The assessment of solvency}.
\bjtitle{ASTIN Bulletin}
\bvolume{6}(\bissue{2}),
\bfpage{79}--\blpage{85}
(\byear{1971})
\doiurl{10.1017/S0515036100010801}
\end{barticle}
\endbibitem

\bibitem[\protect\citeauthoryear{Kassberger
  et~al.}{2008}]{Kassberger2008valuation}
\begin{barticle}
\bauthor{\bsnm{Kassberger}, \binits{S.}},
\bauthor{\bsnm{Kiesel}, \binits{R.}},
\bauthor{\bsnm{Liebmann}, \binits{T.}}:
\batitle{Fair valuation of insurance contracts under lévy process
  specifications}.
\bjtitle{Insurance: Mathematics and Economics}
\bvolume{42}(\bissue{1}),
\bfpage{419}--\blpage{433}
(\byear{2008})
\doiurl{10.1016/j.insmatheco.2007.04.007}
\end{barticle}
\endbibitem

\bibitem[\protect\citeauthoryear{Brodt}{1983}]{Brodt1983MIN}
\begin{barticle}
\bauthor{\bsnm{Brodt}, \binits{A.I.}}:
\batitle{Min-mad life: A multi-period optimization model for life insurance
  company investment decisions}.
\bjtitle{Insurance: Mathematics and Economics}
\bvolume{2}(\bissue{2}),
\bfpage{91}--\blpage{102}
(\byear{1983})
\doiurl{10.1016/0167-6687(83)90017-3}
\end{barticle}
\endbibitem

\bibitem[\protect\citeauthoryear{Brockett and
  Witt}{1982}]{Brockett1982Underwriting}
\begin{barticle}
\bauthor{\bsnm{Brockett}, \binits{P.L.}},
\bauthor{\bsnm{Witt}, \binits{R.C.}}:
\batitle{The underwriting risk and return paradox revisited}.
\bjtitle{The Journal of Risk and Insurance}
\bvolume{49}(\bissue{4}),
\bfpage{621}--\blpage{627}
(\byear{1982})
\doiurl{10.2307/252764}
\end{barticle}
\endbibitem

\bibitem[\protect\citeauthoryear{De~Waegenaere
  et~al.}{2010}]{Waegenaere2010Longevity}
\begin{barticle}
\bauthor{\bsnm{De~Waegenaere}, \binits{A.}},
\bauthor{\bsnm{Melenberg}, \binits{B.}},
\bauthor{\bsnm{Stevens}, \binits{R.}}:
\batitle{Longevity risk}.
\bjtitle{De Economist}
\bvolume{158}(\bissue{2}),
\bfpage{151}--\blpage{192}
(\byear{2010})
\doiurl{10.1007/s10645-010-9143-4}
\end{barticle}
\endbibitem

\bibitem[\protect\citeauthoryear{Ervasti}{2013}]{ervasti2013understanding}
\begin{barticle}
\bauthor{\bsnm{Ervasti}, \binits{M.}}:
\batitle{Understanding and predicting customer behaviour: Framework of value
  dimensions in mobile services}.
\bjtitle{Journal of Customer Behaviour}
\bvolume{12}(\bissue{2-3}),
\bfpage{135}--\blpage{158}
(\byear{2013})
\end{barticle}
\endbibitem

\bibitem[\protect\citeauthoryear{Duffie and Pan}{1997}]{duffie1997overview}
\begin{barticle}
\bauthor{\bsnm{Duffie}, \binits{D.}},
\bauthor{\bsnm{Pan}, \binits{J.}}:
\batitle{An overview of value at risk}.
\bjtitle{Journal of derivatives}
\bvolume{4}(\bissue{3}),
\bfpage{7}--\blpage{49}
(\byear{1997})
\end{barticle}
\endbibitem

\bibitem[\protect\citeauthoryear{Pfaff}{2016}]{pfaff2016financial}
\begin{bbook}
\bauthor{\bsnm{Pfaff}, \binits{B.}}:
\bbtitle{Financial Risk Modelling and Portfolio Optimization with R}.
\bpublisher{John Wiley \& Sons},
\blocation{Hoboken}
(\byear{2016})
\end{bbook}
\endbibitem

\bibitem[\protect\citeauthoryear{Witzany}{2017}]{Witzany2017Credit}
\begin{bbook}
\bauthor{\bsnm{Witzany}, \binits{J.}}:
In: \beditor{\bsnm{Witzany}, \binits{J.}} (ed.)
\bbtitle{Credit Risk Management},
pp. \bfpage{5}--\blpage{18}.
\bpublisher{Springer},
\blocation{Cham}
(\byear{2017}).
\doiurl{10.1007/978-3-319-49800-3_2} .
\burl{https://doi.org/10.1007/978-3-319-49800-3_2}
\end{bbook}
\endbibitem

\bibitem[\protect\citeauthoryear{Moosa}{2007}]{moosa2007operational}
\begin{bbook}
\bauthor{\bsnm{Moosa}, \binits{I.A.}}:
\bbtitle{Operational Risk Management}.
\bpublisher{Springer},
\blocation{Berlin}
(\byear{2007})
\end{bbook}
\endbibitem

\bibitem[\protect\citeauthoryear{Holmström and
  Tirole}{2000}]{Holmström2000Liquidity}
\begin{barticle}
\bauthor{\bsnm{Holmström}, \binits{B.}},
\bauthor{\bsnm{Tirole}, \binits{J.}}:
\batitle{Liquidity and risk management}.
\bjtitle{Journal of Money, Credit and Banking}
\bvolume{32}(\bissue{3}),
\bfpage{295}--\blpage{319}
(\byear{2000})
\doiurl{10.2307/2601167}
\end{barticle}
\endbibitem

\bibitem[\protect\citeauthoryear{Weimer et~al.}{2021}]{rmpWeimer2021Simulation}
\begin{barticle}
\bauthor{\bsnm{Weimer}, \binits{H.}},
\bauthor{\bsnm{Kshetrimayum}, \binits{A.}},
\bauthor{\bsnm{Or\'us}, \binits{R.}}:
\batitle{Simulation methods for open quantum many-body systems}.
\bjtitle{Rev. Mod. Phys.}
\bvolume{93},
\bfpage{015008}
(\byear{2021})
\doiurl{10.1103/RevModPhys.93.015008}
\end{barticle}
\endbibitem

\bibitem[\protect\citeauthoryear{Delgado-Granados
  et~al.}{2024}]{delgado2024quantum}
\begin{botherref}
\oauthor{\bsnm{Delgado-Granados}, \binits{L.H.}},
\oauthor{\bsnm{Krogmeier}, \binits{T.J.}},
\oauthor{\bsnm{Sager-Smith}, \binits{L.M.}},
\oauthor{\bsnm{Avdic}, \binits{I.}},
\oauthor{\bsnm{Hu}, \binits{Z.}},
\oauthor{\bsnm{Sajjan}, \binits{M.}},
\oauthor{\bsnm{Abbasi}, \binits{M.}},
\oauthor{\bsnm{Smart}, \binits{S.E.}},
\oauthor{\bsnm{Narang}, \binits{P.}},
\oauthor{\bsnm{Kais}, \binits{S.}}, et al.:
Quantum algorithms and applications for open quantum systems.
arXiv preprint arXiv:2406.05219
(2024)
\end{botherref}
\endbibitem

\bibitem[\protect\citeauthoryear{Montanaro}{2015}]{montanaro2015quantum}
\begin{barticle}
\bauthor{\bsnm{Montanaro}, \binits{A.}}:
\batitle{Quantum speedup of monte carlo methods}.
\bjtitle{Proceedings of the Royal Society A: Mathematical, Physical and
  Engineering Sciences}
\bvolume{471}(\bissue{2181}),
\bfpage{20150301}
(\byear{2015})
\doiurl{10.1098/rspa.2015.0301} .
\bcomment{doi: 10.1098/rspa.2015.0301}
\end{barticle}
\endbibitem

\bibitem[\protect\citeauthoryear{Layden et~al.}{2023}]{layden2023quantum}
\begin{barticle}
\bauthor{\bsnm{Layden}, \binits{D.}},
\bauthor{\bsnm{Mazzola}, \binits{G.}},
\bauthor{\bsnm{Mishmash}, \binits{R.V.}},
\bauthor{\bsnm{Motta}, \binits{M.}},
\bauthor{\bsnm{Wocjan}, \binits{P.}},
\bauthor{\bsnm{Kim}, \binits{J.-S.}},
\bauthor{\bsnm{Sheldon}, \binits{S.}}:
\batitle{Quantum-enhanced markov chain monte carlo}.
\bjtitle{Nature}
\bvolume{619}(\bissue{7969}),
\bfpage{282}--\blpage{287}
(\byear{2023})
\doiurl{10.1038/s41586-023-06095-4}
\end{barticle}
\endbibitem

\bibitem[\protect\citeauthoryear{Cerezo et~al.}{2021}]{Cerezo2021Variational}
\begin{barticle}
\bauthor{\bsnm{Cerezo}, \binits{M.}},
\bauthor{\bsnm{Arrasmith}, \binits{A.}},
\bauthor{\bsnm{Babbush}, \binits{R.}},
\bauthor{\bsnm{Benjamin}, \binits{S.C.}},
\bauthor{\bsnm{Endo}, \binits{S.}},
\bauthor{\bsnm{Fujii}, \binits{K.}},
\bauthor{\bsnm{McClean}, \binits{J.R.}},
\bauthor{\bsnm{Mitarai}, \binits{K.}},
\bauthor{\bsnm{Yuan}, \binits{X.}},
\bauthor{\bsnm{Cincio}, \binits{L.}}, \betal:
\batitle{Variational quantum algorithms}.
\bjtitle{Nature Reviews Physics}
\bvolume{3}(\bissue{9}),
\bfpage{625}--\blpage{644}
(\byear{2021})
\end{barticle}
\endbibitem

\bibitem[\protect\citeauthoryear{Harrow et~al.}{2009}]{hhl}
\begin{barticle}
\bauthor{\bsnm{Harrow}, \binits{A.W.}},
\bauthor{\bsnm{Hassidim}, \binits{A.}},
\bauthor{\bsnm{Lloyd}, \binits{S.}}:
\batitle{Quantum algorithm for linear systems of equations}.
\bjtitle{Phys. Rev. Lett.}
\bvolume{103},
\bfpage{150502}
(\byear{2009})
\doiurl{10.1103/PhysRevLett.103.150502}
\end{barticle}
\endbibitem

\bibitem[\protect\citeauthoryear{Duan et~al.}{2020}]{Duan2020survey}
\begin{barticle}
\bauthor{\bsnm{Duan}, \binits{B.}},
\bauthor{\bsnm{Yuan}, \binits{J.}},
\bauthor{\bsnm{Yu}, \binits{C.-H.}},
\bauthor{\bsnm{Huang}, \binits{J.}},
\bauthor{\bsnm{Hsieh}, \binits{C.-Y.}}:
\batitle{A survey on hhl algorithm: From theory to application in quantum
  machine learning}.
\bjtitle{Physics Letters A}
\bvolume{384}(\bissue{24}),
\bfpage{126595}
(\byear{2020})
\doiurl{10.1016/j.physleta.2020.126595}
\end{barticle}
\endbibitem

\bibitem[\protect\citeauthoryear{Han et~al.}{2023}]{Han2023vision}
\begin{barticle}
\bauthor{\bsnm{Han}, \binits{K.}},
\bauthor{\bsnm{Wang}, \binits{Y.}},
\bauthor{\bsnm{Chen}, \binits{H.}},
\bauthor{\bsnm{Chen}, \binits{X.}},
\bauthor{\bsnm{Guo}, \binits{J.}},
\bauthor{\bsnm{Liu}, \binits{Z.}},
\bauthor{\bsnm{Tang}, \binits{Y.}},
\bauthor{\bsnm{Xiao}, \binits{A.}},
\bauthor{\bsnm{Xu}, \binits{C.}},
\bauthor{\bsnm{Xu}, \binits{Y.}},
\bauthor{\bsnm{Yang}, \binits{Z.}},
\bauthor{\bsnm{Zhang}, \binits{Y.}},
\bauthor{\bsnm{Tao}, \binits{D.}}:
\batitle{A survey on vision transformer}.
\bjtitle{IEEE Transactions on Pattern Analysis and Machine Intelligence}
\bvolume{45}(\bissue{1}),
\bfpage{87}--\blpage{110}
(\byear{2023})
\doiurl{10.1109/TPAMI.2022.3152247}
\end{barticle}
\endbibitem

\bibitem[\protect\citeauthoryear{Xue
  et~al.}{2024}]{xue2024endtoendquantumvisiontransformer}
\begin{botherref}
\oauthor{\bsnm{Xue}, \binits{C.}},
\oauthor{\bsnm{Chen}, \binits{Z.-Y.}},
\oauthor{\bsnm{Zhuang}, \binits{X.-N.}},
\oauthor{\bsnm{Wang}, \binits{Y.-J.}},
\oauthor{\bsnm{Sun}, \binits{T.-P.}},
\oauthor{\bsnm{Wang}, \binits{J.-C.}},
\oauthor{\bsnm{Liu}, \binits{H.-Y.}},
\oauthor{\bsnm{Wu}, \binits{Y.-C.}},
\oauthor{\bsnm{Wang}, \binits{Z.-L.}},
\oauthor{\bsnm{Guo}, \binits{G.-P.}}:
End-to-End Quantum Vision Transformer: Towards Practical Quantum Speedup in
  Large-Scale Models
(2024).
\url{https://arxiv.org/abs/2402.18940}
\end{botherref}
\endbibitem

\bibitem[\protect\citeauthoryear{Bravo-Prieto et~al.}{2020}]{Bravo2020quantum}
\begin{barticle}
\bauthor{\bsnm{Bravo-Prieto}, \binits{C.}},
\bauthor{\bsnm{Garc\'{\i}a-Mart\'{\i}n}, \binits{D.}},
\bauthor{\bsnm{Latorre}, \binits{J.I.}}:
\batitle{Quantum singular value decomposer}.
\bjtitle{Phys. Rev. A}
\bvolume{101},
\bfpage{062310}
(\byear{2020})
\doiurl{10.1103/PhysRevA.101.062310}
\end{barticle}
\endbibitem

\bibitem[\protect\citeauthoryear{Albash and Lidar}{2018}]{albash2018adiabatic}
\begin{barticle}
\bauthor{\bsnm{Albash}, \binits{T.}},
\bauthor{\bsnm{Lidar}, \binits{D.A.}}:
\batitle{Adiabatic quantum computation}.
\bjtitle{Rev. Mod. Phys.}
\bvolume{90},
\bfpage{015002}
(\byear{2018})
\doiurl{10.1103/RevModPhys.90.015002}
\end{barticle}
\endbibitem

\bibitem[\protect\citeauthoryear{Rosenberg et~al.}{2015}]{rosenberg2015solving}
\begin{bchapter}
\bauthor{\bsnm{Rosenberg}, \binits{G.}},
\bauthor{\bsnm{Haghnegahdar}, \binits{P.}},
\bauthor{\bsnm{Goddard}, \binits{P.}},
\bauthor{\bsnm{Carr}, \binits{P.}},
\bauthor{\bsnm{Wu}, \binits{K.}},
\bauthor{\bsnm{Prado}, \binits{M.L.}}:
\bctitle{Solving the optimal trading trajectory problem using a quantum
  annealer}.
\bsertitle{WHPCF '15}.
\bpublisher{Association for Computing Machinery},
\blocation{New York, NY, USA}
(\byear{2015}).
\doiurl{10.1145/2830556.2830563} .
\burl{https://doi.org/10.1145/2830556.2830563}
\end{bchapter}
\endbibitem

\bibitem[\protect\citeauthoryear{Venturelli and
  Kondratyev}{2019}]{venturelli2019reverse}
\begin{barticle}
\bauthor{\bsnm{Venturelli}, \binits{D.}},
\bauthor{\bsnm{Kondratyev}, \binits{A.}}:
\batitle{Reverse quantum annealing approach to portfolio optimization
  problems}.
\bjtitle{Quantum Machine Intelligence}
\bvolume{1}(\bissue{1}),
\bfpage{17}--\blpage{30}
(\byear{2019})
\doiurl{10.1007/s42484-019-00001-w}
\end{barticle}
\endbibitem

\bibitem[\protect\citeauthoryear{Farhi et~al.}{2014}]{QAOA}
\begin{botherref}
\oauthor{\bsnm{Farhi}, \binits{E.}},
\oauthor{\bsnm{Goldstone}, \binits{J.}},
\oauthor{\bsnm{Gutmann}, \binits{S.}}:
A Quantum Approximate Optimization Algorithm
(2014).
\url{https://arxiv.org/abs/1411.4028}
\end{botherref}
\endbibitem

\bibitem[\protect\citeauthoryear{Dou et~al.}{2022}]{qpanda}
\begin{botherref}
\oauthor{\bsnm{Dou}, \binits{M.}},
\oauthor{\bsnm{Zou}, \binits{T.}},
\oauthor{\bsnm{Fang}, \binits{Y.}},
\oauthor{\bsnm{Wang}, \binits{J.}},
\oauthor{\bsnm{Zhao}, \binits{D.}},
\oauthor{\bsnm{Yu}, \binits{L.}},
\oauthor{\bsnm{Chen}, \binits{B.}},
\oauthor{\bsnm{Guo}, \binits{W.}},
\oauthor{\bsnm{Li}, \binits{Y.}},
\oauthor{\bsnm{Chen}, \binits{Z.}},
\oauthor{\bsnm{Guo}, \binits{G.}}:
QPanda: high-performance quantum computing framework for multiple application
  scenarios
(2022).
\url{https://arxiv.org/abs/2212.14201}
\end{botherref}
\endbibitem

\bibitem[\protect\citeauthoryear{Mahul and Stutley}{2010}]{mahul2010government}
\begin{bbook}
\bauthor{\bsnm{Mahul}, \binits{O.}},
\bauthor{\bsnm{Stutley}, \binits{C.J.}}:
\bbtitle{Government Support to Agricultural Insurance: Challenges and Options
  for Developing Countries}.
\bpublisher{World Bank Publications}, \blocation{???}
(\byear{2010})
\end{bbook}
\endbibitem

\bibitem[\protect\citeauthoryear{Xu et~al.}{2010}]{xu2010systemic}
\begin{barticle}
\bauthor{\bsnm{Xu}, \binits{W.}},
\bauthor{\bsnm{Filler}, \binits{G.}},
\bauthor{\bsnm{Odening}, \binits{M.}},
\bauthor{\bsnm{Okhrin}, \binits{O.}}:
\batitle{On the systemic nature of weather risk}.
\bjtitle{Agricultural Finance Review}
\bvolume{70}(\bissue{2}),
\bfpage{267}--\blpage{284}
(\byear{2010})
\doiurl{10.1108/00021461011065283}
\end{barticle}
\endbibitem

\bibitem[\protect\citeauthoryear{Miranda and
  Glauber}{1997}]{miranda1997systemic}
\begin{barticle}
\bauthor{\bsnm{Miranda}, \binits{M.J.}},
\bauthor{\bsnm{Glauber}, \binits{J.W.}}:
\batitle{Systemic risk, reinsurance, and the failure of crop insurance
  markets}.
\bjtitle{American Journal of Agricultural Economics}
\bvolume{79}(\bissue{1}),
\bfpage{206}--\blpage{215}
(\byear{1997})
\doiurl{10.2307/1243954}
\end{barticle}
\endbibitem

\bibitem[\protect\citeauthoryear{Porth et~al.}{2015}]{porth2015portfolio}
\begin{barticle}
\bauthor{\bsnm{Porth}, \binits{L.}},
\bauthor{\bsnm{Pai}, \binits{J.}},
\bauthor{\bsnm{Boyd}, \binits{M.}}:
\batitle{A portfolio optimization approach using combinatorics with a genetic
  algorithm for developing a reinsurance model}.
\bjtitle{Journal of Risk and Insurance}
\bvolume{82}(\bissue{3}),
\bfpage{687}--\blpage{713}
(\byear{2015})
\doiurl{10.1111/jori.12037}
\end{barticle}
\endbibitem

\bibitem[\protect\citeauthoryear{Kerenidis
  et~al.}{2022}]{kerenidis2021classical}
\begin{botherref}
\oauthor{\bsnm{Kerenidis}, \binits{I.}},
\oauthor{\bsnm{Landman}, \binits{J.}},
\oauthor{\bsnm{Mathur}, \binits{N.}}:
Classical and quantum algorithms for orthogonal neural networks
(2022)
{\href{https://arxiv.org/abs/2106.07198}{{arXiv:2106.07198}}}
{[quant-ph]}
\end{botherref}
\endbibitem

\bibitem[\protect\citeauthoryear{Johri et~al.}{2021}]{johri2021nearest}
\begin{barticle}
\bauthor{\bsnm{Johri}, \binits{S.}},
\bauthor{\bsnm{Debnath}, \binits{S.}},
\bauthor{\bsnm{Mocherla}, \binits{A.}},
\bauthor{\bsnm{Singk}, \binits{A.}},
\bauthor{\bsnm{Prakash}, \binits{A.}},
\bauthor{\bsnm{Kim}, \binits{J.}},
\bauthor{\bsnm{Kerenidis}, \binits{I.}}:
\batitle{Nearest centroid classification on a trapped ion quantum computer}.
\bjtitle{npj Quantum Information}
\bvolume{7}(\bissue{1}),
\bfpage{122}
(\byear{2021})
\doiurl{10.1038/s41534-021-00456-5}
\end{barticle}
\endbibitem

\bibitem[\protect\citeauthoryear{Yuan et~al.}{2024}]{rbsrev}
\begin{botherref}
\oauthor{\bsnm{Yuan}, \binits{H.}},
\oauthor{\bsnm{Long}, \binits{C.K.}},
\oauthor{\bsnm{Lepage}, \binits{H.V.}},
\oauthor{\bsnm{Barnes}, \binits{C.H.W.}}:
Quantifying the advantages of applying quantum approximate algorithms to
  portfolio optimisation
(2024).
\url{https://arxiv.org/abs/2410.16265}
\end{botherref}
\endbibitem

\bibitem[\protect\citeauthoryear{Arute et~al.}{2019}]{arute2019quantum}
\begin{barticle}
\bauthor{\bsnm{Arute}, \binits{F.}},
\bauthor{\bsnm{Arya}, \binits{K.}},
\bauthor{\bsnm{Babbush}, \binits{R.}},
\bauthor{\bsnm{Bacon}, \binits{D.}},
\bauthor{\bsnm{Bardin}, \binits{J.C.}},
\bauthor{\bsnm{Barends}, \binits{R.}},
\bauthor{\bsnm{Biswas}, \binits{R.}},
\bauthor{\bsnm{Boixo}, \binits{S.}},
\bauthor{\bsnm{Brandao}, \binits{F.G.S.L.}},
\bauthor{\bsnm{Buell}, \binits{D.A.}},
\bauthor{\bsnm{Burkett}, \binits{B.}},
\bauthor{\bsnm{Chen}, \binits{Y.}},
\bauthor{\bsnm{Chen}, \binits{Z.}},
\bauthor{\bsnm{Chiaro}, \binits{B.}},
\bauthor{\bsnm{Collins}, \binits{R.}},
\bauthor{\bsnm{Courtney}, \binits{W.}},
\bauthor{\bsnm{Dunsworth}, \binits{A.}},
\bauthor{\bsnm{Farhi}, \binits{E.}},
\bauthor{\bsnm{Foxen}, \binits{B.}},
\bauthor{\bsnm{Fowler}, \binits{A.}},
\bauthor{\bsnm{Gidney}, \binits{C.}},
\bauthor{\bsnm{Giustina}, \binits{M.}},
\bauthor{\bsnm{Graff}, \binits{R.}},
\bauthor{\bsnm{Guerin}, \binits{K.}},
\bauthor{\bsnm{Habegger}, \binits{S.}},
\bauthor{\bsnm{Harrigan}, \binits{M.P.}},
\bauthor{\bsnm{Hartmann}, \binits{M.J.}},
\bauthor{\bsnm{Ho}, \binits{A.}},
\bauthor{\bsnm{Hoffmann}, \binits{M.}},
\bauthor{\bsnm{Huang}, \binits{T.}},
\bauthor{\bsnm{Humble}, \binits{T.S.}},
\bauthor{\bsnm{Isakov}, \binits{S.V.}},
\bauthor{\bsnm{Jeffrey}, \binits{E.}},
\bauthor{\bsnm{Jiang}, \binits{Z.}},
\bauthor{\bsnm{Kafri}, \binits{D.}},
\bauthor{\bsnm{Kechedzhi}, \binits{K.}},
\bauthor{\bsnm{Kelly}, \binits{J.}},
\bauthor{\bsnm{Klimov}, \binits{P.V.}},
\bauthor{\bsnm{Knysh}, \binits{S.}},
\bauthor{\bsnm{Korotkov}, \binits{A.}},
\bauthor{\bsnm{Kostritsa}, \binits{F.}},
\bauthor{\bsnm{Landhuis}, \binits{D.}},
\bauthor{\bsnm{Lindmark}, \binits{M.}},
\bauthor{\bsnm{Lucero}, \binits{E.}},
\bauthor{\bsnm{Lyakh}, \binits{D.}},
\bauthor{\bsnm{Mandrà}, \binits{S.}},
\bauthor{\bsnm{McClean}, \binits{J.R.}},
\bauthor{\bsnm{McEwen}, \binits{M.}},
\bauthor{\bsnm{Megrant}, \binits{A.}},
\bauthor{\bsnm{Mi}, \binits{X.}},
\bauthor{\bsnm{Michielsen}, \binits{K.}},
\bauthor{\bsnm{Mohseni}, \binits{M.}},
\bauthor{\bsnm{Mutus}, \binits{J.}},
\bauthor{\bsnm{Naaman}, \binits{O.}},
\bauthor{\bsnm{Neeley}, \binits{M.}},
\bauthor{\bsnm{Neill}, \binits{C.}},
\bauthor{\bsnm{Niu}, \binits{M.Y.}},
\bauthor{\bsnm{Ostby}, \binits{E.}},
\bauthor{\bsnm{Petukhov}, \binits{A.}},
\bauthor{\bsnm{Platt}, \binits{J.C.}},
\bauthor{\bsnm{Quintana}, \binits{C.}},
\bauthor{\bsnm{Rieffel}, \binits{E.G.}},
\bauthor{\bsnm{Roushan}, \binits{P.}},
\bauthor{\bsnm{Rubin}, \binits{N.C.}},
\bauthor{\bsnm{Sank}, \binits{D.}},
\bauthor{\bsnm{Satzinger}, \binits{K.J.}},
\bauthor{\bsnm{Smelyanskiy}, \binits{V.}},
\bauthor{\bsnm{Sung}, \binits{K.J.}},
\bauthor{\bsnm{Trevithick}, \binits{M.D.}},
\bauthor{\bsnm{Vainsencher}, \binits{A.}},
\bauthor{\bsnm{Villalonga}, \binits{B.}},
\bauthor{\bsnm{White}, \binits{T.}},
\bauthor{\bsnm{Yao}, \binits{Z.J.}},
\bauthor{\bsnm{Yeh}, \binits{P.}},
\bauthor{\bsnm{Zalcman}, \binits{A.}},
\bauthor{\bsnm{Neven}, \binits{H.}},
\bauthor{\bsnm{Martinis}, \binits{J.M.}}:
\batitle{Quantum supremacy using a programmable superconducting processor}.
\bjtitle{Nature}
\bvolume{574}(\bissue{7779}),
\bfpage{505}--\blpage{510}
(\byear{2019})
\doiurl{10.1038/s41586-019-1666-5}
\end{barticle}
\endbibitem

\bibitem[\protect\citeauthoryear{Gambetta et~al.}{2017}]{Gambetta_2017}
\begin{barticle}
\bauthor{\bsnm{Gambetta}, \binits{J.M.}},
\bauthor{\bsnm{Chow}, \binits{J.M.}},
\bauthor{\bsnm{Steffen}, \binits{M.}}:
\batitle{Building logical qubits in a superconducting quantum computing
  system}.
\bjtitle{npj Quantum Information}
\bvolume{3}(\bissue{1}),
\bfpage{2}
(\byear{2017})
\doiurl{10.1038/s41534-016-0004-0}
\end{barticle}
\endbibitem

\bibitem[\protect\citeauthoryear{Campbell et~al.}{2017}]{Campbell_2017}
\begin{barticle}
\bauthor{\bsnm{Campbell}, \binits{E.T.}},
\bauthor{\bsnm{Terhal}, \binits{B.M.}},
\bauthor{\bsnm{Vuillot}, \binits{C.}}:
\batitle{Roads towards fault-tolerant universal quantum computation}.
\bjtitle{Nature}
\bvolume{549}(\bissue{7671}),
\bfpage{172}--\blpage{179}
(\byear{2017})
\doiurl{10.1038/nature23460}
\end{barticle}
\endbibitem

\bibitem[\protect\citeauthoryear{Coecke et~al.}{2020}]{coecke2020foundations}
\begin{botherref}
\oauthor{\bsnm{Coecke}, \binits{B.}},
\oauthor{\bsnm{Felice}, \binits{G.}},
\oauthor{\bsnm{Meichanetzidis}, \binits{K.}},
\oauthor{\bsnm{Toumi}, \binits{A.}}:
Foundations for near-term quantum natural language processing
(2020)
{\href{https://arxiv.org/abs/2012.03755}{{arXiv:2012.03755}}}
{[quant-ph]}
\end{botherref}
\endbibitem

\bibitem[\protect\citeauthoryear{Trugenberger}{2002}]{trugenberger2002quantum}
\begin{barticle}
\bauthor{\bsnm{Trugenberger}, \binits{C.A.}}:
\batitle{Quantum pattern recognition}.
\bjtitle{Quantum Information Processing}
\bvolume{1}(\bissue{6}),
\bfpage{471}--\blpage{493}
(\byear{2002})
\doiurl{10.1023/A:1024022632303}
\end{barticle}
\endbibitem

\bibitem[\protect\citeauthoryear{Zhang et~al.}{2022}]{qsp1}
\begin{barticle}
\bauthor{\bsnm{Zhang}, \binits{X.-M.}},
\bauthor{\bsnm{Li}, \binits{T.}},
\bauthor{\bsnm{Yuan}, \binits{X.}}:
\batitle{Quantum state preparation with optimal circuit depth: Implementations
  and applications}.
\bjtitle{Phys. Rev. Lett.}
\bvolume{129},
\bfpage{230504}
(\byear{2022})
\doiurl{10.1103/PhysRevLett.129.230504}
\end{barticle}
\endbibitem

\bibitem[\protect\citeauthoryear{Sun et~al.}{2023}]{qsp2}
\begin{barticle}
\bauthor{\bsnm{Sun}, \binits{X.}},
\bauthor{\bsnm{Tian}, \binits{G.}},
\bauthor{\bsnm{Yang}, \binits{S.}},
\bauthor{\bsnm{Yuan}, \binits{P.}},
\bauthor{\bsnm{Zhang}, \binits{S.}}:
\batitle{Asymptotically optimal circuit depth for quantum state preparation and
  general unitary synthesis}.
\bjtitle{IEEE Transactions on Computer-Aided Design of Integrated Circuits and
  Systems}
\bvolume{42}(\bissue{10}),
\bfpage{3301}--\blpage{3314}
(\byear{2023})
\doiurl{10.1109/TCAD.2023.3244885}
\end{barticle}
\endbibitem

\bibitem[\protect\citeauthoryear{Giovannetti
  et~al.}{2008}]{Giovannetti2008qram}
\begin{barticle}
\bauthor{\bsnm{Giovannetti}, \binits{V.}},
\bauthor{\bsnm{Lloyd}, \binits{S.}},
\bauthor{\bsnm{Maccone}, \binits{L.}}:
\batitle{Quantum random access memory}.
\bjtitle{Phys. Rev. Lett.}
\bvolume{100},
\bfpage{160501}
(\byear{2008})
\doiurl{10.1103/PhysRevLett.100.160501}
\end{barticle}
\endbibitem

\bibitem[\protect\citeauthoryear{Zhuang et~al.}{2024}]{sipqc}
\begin{botherref}
\oauthor{\bsnm{Zhuang}, \binits{X.-N.}},
\oauthor{\bsnm{Chen}, \binits{Z.-Y.}},
\oauthor{\bsnm{Xue}, \binits{C.}},
\oauthor{\bsnm{Xu}, \binits{X.-F.}},
\oauthor{\bsnm{Wang}, \binits{C.}},
\oauthor{\bsnm{Liu}, \binits{H.-Y.}},
\oauthor{\bsnm{Sun}, \binits{T.-P.}},
\oauthor{\bsnm{Wang}, \binits{Y.-J.}},
\oauthor{\bsnm{Wu}, \binits{Y.-C.}},
\oauthor{\bsnm{Guo}, \binits{G.-P.}}:
Statistics-Informed Parameterized Quantum Circuit via Maximum Entropy Principle
  for Data Science and Finance
(2024).
\url{https://arxiv.org/abs/2406.01335}
\end{botherref}
\endbibitem

\bibitem[\protect\citeauthoryear{Aaronson}{2018}]{st}
\begin{bchapter}
\bauthor{\bsnm{Aaronson}, \binits{S.}}:
\bctitle{Shadow tomography of quantum states}.
\bsertitle{STOC 2018},
pp. \bfpage{325}--\blpage{338}.
\bpublisher{Association for Computing Machinery},
\blocation{New York, NY, USA}
(\byear{2018}).
\doiurl{10.1145/3188745.3188802} .
\burl{https://doi.org/10.1145/3188745.3188802}
\end{bchapter}
\endbibitem

\bibitem[\protect\citeauthoryear{Huang et~al.}{2020}]{st2}
\begin{barticle}
\bauthor{\bsnm{Huang}, \binits{H.-Y.}},
\bauthor{\bsnm{Kueng}, \binits{R.}},
\bauthor{\bsnm{Preskill}, \binits{J.}}:
\batitle{Predicting many properties of a quantum system from very few
  measurements}.
\bjtitle{Nature Physics}
\bvolume{16}(\bissue{10}),
\bfpage{1050}--\blpage{1057}
(\byear{2020})
\doiurl{10.1038/s41567-020-0932-7}
\end{barticle}
\endbibitem

\bibitem[\protect\citeauthoryear{Cai et~al.}{2023}]{qem}
\begin{barticle}
\bauthor{\bsnm{Cai}, \binits{Z.}},
\bauthor{\bsnm{Babbush}, \binits{R.}},
\bauthor{\bsnm{Benjamin}, \binits{S.C.}},
\bauthor{\bsnm{Endo}, \binits{S.}},
\bauthor{\bsnm{Huggins}, \binits{W.J.}},
\bauthor{\bsnm{Li}, \binits{Y.}},
\bauthor{\bsnm{McClean}, \binits{J.R.}},
\bauthor{\bsnm{O'Brien}, \binits{T.E.}}:
\batitle{Quantum error mitigation}.
\bjtitle{Rev. Mod. Phys.}
\bvolume{95},
\bfpage{045005}
(\byear{2023})
\doiurl{10.1103/RevModPhys.95.045005}
\end{barticle}
\endbibitem

\bibitem[\protect\citeauthoryear{Ruiz-Perez and
  Garcia-Escartin}{2017}]{Ruiz_Perez_2017}
\begin{barticle}
\bauthor{\bsnm{Ruiz-Perez}, \binits{L.}},
\bauthor{\bsnm{Garcia-Escartin}, \binits{J.C.}}:
\batitle{Quantum arithmetic with the quantum fourier transform}.
\bjtitle{Quantum Information Processing}
\bvolume{16}(\bissue{6}),
\bfpage{152}
(\byear{2017})
\doiurl{10.1007/s11128-017-1603-1}
\end{barticle}
\endbibitem

\end{thebibliography}
\end{document}